\documentclass[%
 reprint,
 amsmath,amssymb,
 aps,
prfluids
]{revtex4-2}

\usepackage{url}

\usepackage{graphicx}
\usepackage{dcolumn}
\usepackage{bm}

\usepackage{enumitem}
\setlist[itemize]{itemsep=0.2em, topsep=0pt, parsep=0pt, partopsep=0pt}

\usepackage{tikz}
\usetikzlibrary{arrows.meta,positioning,calc}

\newcommand{\abscite}[1]{\,[#1]}

\usepackage{subcaption}

\begin{document}

\preprint{APS/123-QED}

\title{Order of Magnitude Analysis and Data-Based\\
Physics-Informed Symbolic Regression for Turbulent Pipe Flow}

\author{Yunus Emre \"Unal}
\affiliation{Fluid Dynamics and Spray Laboratory, Mechanical Engineering Department, \"Ozyegin University, Istanbul, T\"urkiye}

\author{\"Ozg\"ur Ertun\c{c}}
\email{ozgur.ertunc@ozyegin.edu.tr}
\affiliation{Fluid Dynamics and Spray Laboratory, Mechanical Engineering Department, \"Ozyegin University, Istanbul, T\"urkiye}

\author{Ismail Ari}
\affiliation{Cloud Computing Research Laboratory, Computer Science Department, \"Ozyegin University, Istanbul, T\"urkiye}

\author{Ivan Oti\'c}
\affiliation{Institute for Thermal Energy Technology and Safety, Karlsruhe Institute of Technology, Karlsruhe, Germany}

\date{\today}





\begin{abstract}
Friction losses in rough pipes are commonly predicted with semi-empirical correlations
such as the Colebrook--White equation \abscite{C. F. Colebrook, J. Inst. Civ. Eng. 11, 133 (1939)} and explicit approximations (e.g.\ Haaland) \abscite{S. E. Haaland, J. Fluids Eng. 105, 89 (1983)},
which interpolate decades of data but do not fully reproduce the behaviour of
Nikuradse’s rough pipe experiments \abscite{J. Nikuradse, NACA Tech. Memo. 1292 (1950)}. Here, we derive scaling relations for the
viscous and turbulent contributions to the streamwise pressure drop using an order-of-magnitude analysis (OMA) of the Reynolds-averaged
Navier-Stokes equations and the mean/turbulent kinetic-energy transport. These
relations yield four quantitative physical constraints that bound the local
sensitivities of $\Delta P$ to mean velocity, roughness, viscosity and density
through exponent envelopes, and serve as a physics prior for symbolic
regression. Using the combined data of Nikuradse and smooth-pipe measurements \abscite{M. V. Zagarola and A. J. Smits, Phys. Rev. Lett. 80, 239 (1998)}, we seek compact and explicit correlations for
$f=g(\mathrm{Re},\varepsilon/D)$ that fit the experiments while respecting the
OMA constraints and recovering correct smooth and fully rough asymptotics rather than simply tuning another empirical interpolation between smooth-wall and fully rough limits. We modify a multigene genetic-programming engine (GPTIPS2) to jointly optimize tree structure, linear gene weights  and embedded constants, and evaluate each model by three minimization objectives: fitness, structural complexity and a physics score measuring constraint violation; a three-dimensional Pareto front is evolved rather than a single regularized loss. The method discovers short,
interpretable expressions that satisfy the constraints while accurately reproducing friction factors across smooth-to-rough and over a wide Reynolds number range. The best fitting correlation is validated on rough, high-$Re$ campaigns \abscite{C. J. Swanson \textit{et al.}, J. Fluid Mech. 461, 51 (2002); M. A. Shockling \textit{et al.}, J. Fluid Mech. 564, 267 (2006); L. I. Langelandsvik \textit{et al.}, J. Fluid Mech. 595, 323 (2008)} up to
$Re\sim 10^7$.
The framework generalizes to other nondimensional correlation problems where only leading-order balances are known.
\end{abstract}

\maketitle

%

\section{\label{sec:intro}Introduction}


Predictive correlations for frictional pressure loss in pipe flows are central to
hydraulic design, flow metering, and energy systems.  Classical treatments express
the Darcy--Weisbach friction factor \(f\) as a function of the Reynolds number
\(Re\) and the relative roughness \(\varepsilon/D\), and summarize decades of
experiments in compact formulas such as the Colebrook–White equation
and its explicit approximations such as Haaland equation, together with the Moody diagram that engineers
use in practice \citep{Colebrook1939,Moody1944,Haaland1983,White2006}.  The
underlying structure is inherently dimensionless: once the flow is fully developed,
the pressure gradient can depend only on the ratios of inertial to viscous forces
and on the relative wall-roughness.

Dimensional analysis provides a natural starting point for such problems.
Buckingham-\(\Pi\) theorem guarantees that any physically admissible relation among
\(\Delta P\), \(\overline{U}_m\), \(\rho\), \(\mu\), \(D\), \(L\), and
\(\varepsilon\) may be written in terms of a reduced set of dimensionless groups
\citep{Buckingham1914,Bridgman1922}.  For internal flows, this leads to the familiar
choice of the Reynolds number and the relative roughness, so that
\(f = f(Re,\varepsilon/D)\).  However, the theorem is silent on which
dimensionless groups are most predictive and on how they should be combined.
Historically, this ranking has been supplied by physical reasoning and experiments:
Blasius used smooth-pipe data to infer a nearly power-law dependence of the friction
factor on \(Re\) \citep{Blasius1913}, while Nikuradse's sand-grain roughness
experiments mapped out the dependence on
\(\varepsilon/D\) and extended the data into the fully rough regime
\citep{Nikuradse1950}. At the same time, it has long been recognized that roughness morphology matters: compared with Nikuradse's uniform sand grains, commercial pipes with broader distributions of protuberance sizes can exhibit a more gradual departure from the smooth-pipe law. In that setting, \(\varepsilon\) is most consistently interpreted as an effective (equivalent sand-grain) roughness referenced to Nikuradse-style behaviour, enabling different surface types to be compared on a common roughness scale \citep{Colebrook1937}. In a similar spirit, earlier resistance formulations in hydraulics also emphasized relative roughness measures and reported weak power dependence of roughness coefficients on an absolute roughness scale \citep{Strickler1981}. Semi-empirical correlations such as
Colebrook–White interpolate between these regimes but remain implicit and
non-trivial to invert.

Recent years have seen growing interest in using data-driven methods to discover
dimensionless groups and closure relations directly from experimental or numerical
data.  The "Virtual Nikuradse" model, for example, constructs a high-fidelity
friction-factor surface from Nikuradse-style data via spline and polynomial fits
\citep{Yang2009}.  More broadly, symbolic regression and related techniques
seek analytical expressions that reproduce observed data while remaining compact and
interpretable \citep{Koza1992,Schmidt2009,Udrescu2020,Cranmer2020}.
The GPTIPS 2 framework, in particular, provides a
multigene genetic-programming implementation for symbolic regression in MATLAB
\citep{Searson2015}, and tools such as PySR offer efficient modern solvers in
Python/Julia \citep{Cranmer2020}.  Yet most of these applications only enforce
dimensional homogeneity and perhaps simple bounds or penalties; they typically do
not encode the detailed asymptotics and monotonicity properties that are known for
pipe friction.
As a result, purely data-driven symbolic models may match friction-factor data over
the training range but exhibit unphysical behaviour when extrapolated: exponents in
\(\Delta P\) versus velocity may drift outside plausible ranges, pressure drop may
decrease with increasing roughness in some corner of parameter space, or the
dependence on viscosity and density may contradict basic momentum-balance
considerations.  For engineering use, such pathologies are unacceptable even if they occur in regions with sparse data.

In this work, we combine a problem-specific order-of-magnitude analysis (OMA) with physics-constrained symbolic regression to discover new correlations for the friction factor in turbulent pipe flows. Starting from
a stream-wise momentum balance and energy equations in the framework of Reynolds decomposition and averaging, we derive an OMA model for the pressure
drop that has separate viscous and roughness-dominated turbulent contributions. The two contributions are blended with a logistic function so that the resulting equation can reflect the transition from viscous effects-dominated smooth turbulent pipe flow to inertial turbulence-dominated rough turbulent pipe flow.  Written in dimensionless form, this OMA model implies specific, testable trends for how the pressure drop behaves when one physical parameter is varied while the others are held fixed. We characterize these trends through four logarithmic sensitivities:
(i) the effective velocity exponent,
(ii) the roughness sensitivity,
(iii) the viscous sensitivity, and
(iv) the density sensitivity.
Evaluated over the Nikuradse and Superpipe parameter ranges, the OMA model yields
tight envelopes for these quantities.

We then embed these constraints in a modified symbolic-regression pipeline built on
GPTIPS 2.  The input space consists of a grid of candidate dimensionless groups
generated by Buckingham-\(\Pi\) analysis; individuals are multi-gene expressions that
linearly combine symbolic subtrees of these groups.  For each candidate model, we
first refit all linear gene weights and embedded constants by a two-stage procedure:
ridge-regularized least squares followed by joint optimization of all constants via
Nelder–Mead, using a fast evaluation of the OMA-inspired physics score.  The final
model is evaluated on three objectives:
(i) fitness (normalised RMSE of \(f\)),
(ii) a structural complexity measure, and
(iii) a physics score that summarizes the
worst violation of the OMA constraints.  Multi-objective selection with Pareto
sorting and crowding distance is applied directly to these three objective scores rather than folding physics
into a single penalty term, so that accuracy, simplicity, and physical consistency
can be traded explicitly.

The contributions of this paper are threefold.  First, we derive an OMA model for
pressure drop in rough turbulent pipes that yields quantitative, data-calibrated
constraints on the local power-law exponents with respect to velocity, roughness,
viscosity, and density.  Second, we design a physics-informed symbolic-regression
workflow that augments GPTIPS 2 with joint constant optimization and a
three-objective Pareto search driven by these OMA constraints.  Third, applying this
framework to  Nikuradse and Superpipe datasets, we obtain a compact explicit
correlations for \(\Delta P\) and \(f(Re,\varepsilon/D)\) that match or exceed the
accuracy of classical formulas while remaining consistent with the asymptotic
behaviour implied by the OMA analysis.
\section{Methodology}
\subsection{Derivation of Governing Equations}
\begin{figure}
  \centering

  \begin{subfigure}{\columnwidth}
    \centering
    \includegraphics[width=0.9\columnwidth]{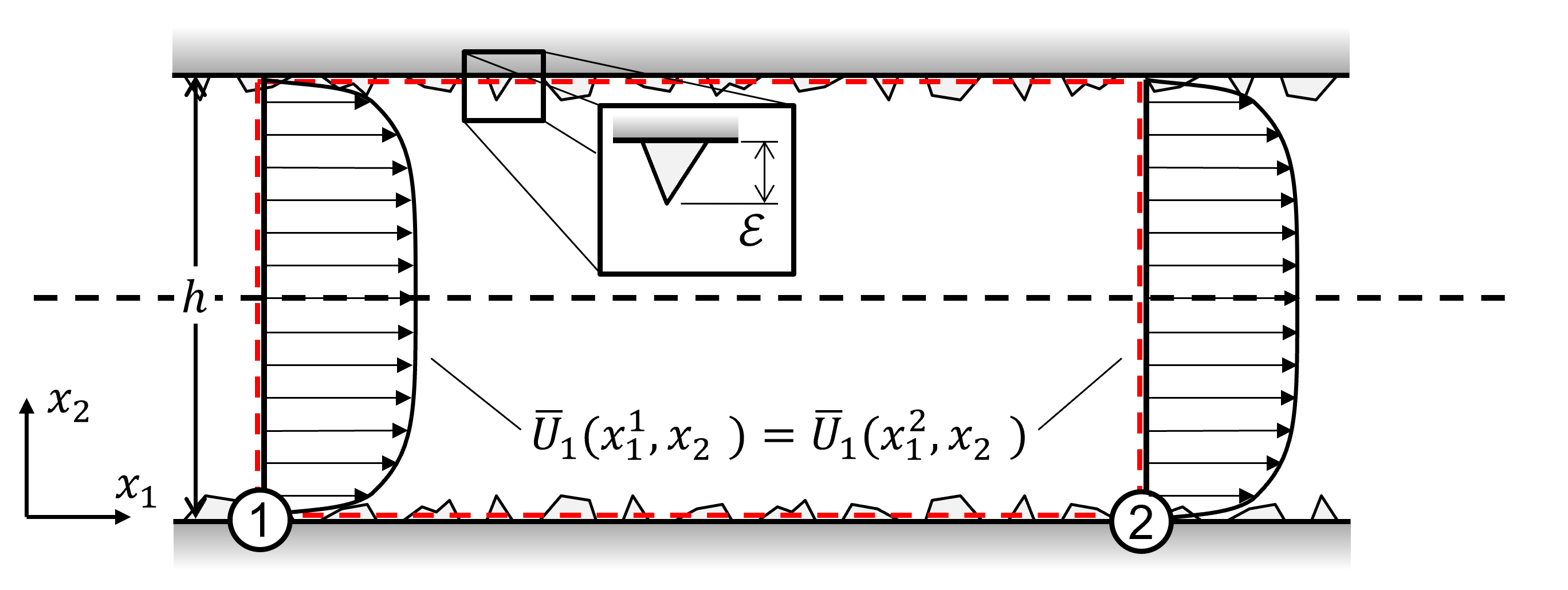}
    \caption{}
    \label{fig:CVa}
  \end{subfigure}

  \vspace{0.6em}

  \begin{subfigure}{\columnwidth}
    \centering
    \includegraphics[width=0.9\columnwidth]{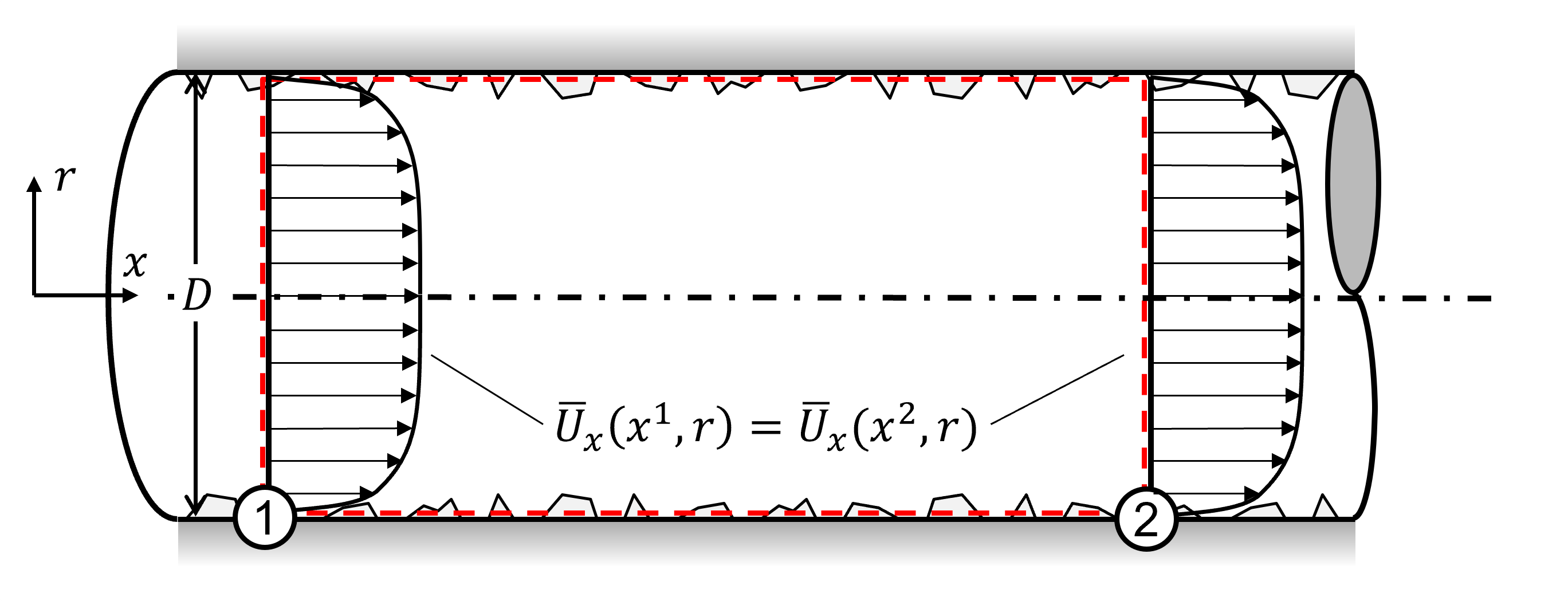}
    \caption{}
    \label{fig:CVb}
  \end{subfigure}

  \caption{Control volume (CV, red dashed region) for an incompressible fully developed turbulent statistically stationary (a) two-dimensional channel and (b) pipe flow.}
  \label{fig:CV}
\end{figure}

The symbolic regression process is, in principle, an optimization 
process, and the constraints of this process are derived from the order-of-magnitude analysis. The Reynolds transport equations, along with Reynolds decomposition and averaging, provide a suitable framework for turbulent flows. Without delving into the complexity of cylindrical coordinates, the analysis is conducted for an incompressible, fully developed horizontal channel flow. A control volume (CV) shown in  Fig.~\ref{fig:CVa} is selected from the fully developed portion of the channel having a height of $h$, and a width of $w$. The channel portion has a length of $L$ and a width much longer than its height ($w>>h$), so the flow is assumed to be two-dimensional. Hence, the momentum equation for this control volume can be written as:
\begin{equation}
    \Sigma F_i =\frac{\partial}{\partial t} \int_{CV} \rho U_i \, dV + 
    \int_{CS} \rho U_i U_j n_j \, dA
    \label{eq:reynolds_transport_momentum}
\end{equation}
where $\Sigma F_i$ is the net external force in the $i$-direction acting on the control volume, $\rho$ is the fluid density, $U_i$ is the $i$th component of velocity, $n_j$ is the outward unit normal to the control surface, and $dV$ and $ dA$ denote differential volume and area elements, respectively; repeated indices imply summation.


The second term on the right-hand side (RHS) is the net flow rate of momentum, whereas the first term on the RHS is the rate of change of momentum in CV. In addition, the surface integral can be transformed into a volume integral by the divergence theorem:
\begin{equation}
    \Sigma F_i =\frac{\partial}{\partial t} \int_{CV} \rho U_i \, dV + 
    \int_{CV} \frac{\partial}{\partial x_j} \rho U_i U_j \, dV
    \label{eq:reynolds_transport_momentum_CV}
\end{equation}
To tackle the turbulence, the Reynolds decomposition is used, and all the variables, except $\rho$, can be written as the sum of the ensemble average and the fluctuating component: 
\begin{align}
    \Sigma (\overline{F}_i+f_i) =& \frac{\partial}{\partial t} \int_{CV} \rho (\overline{U}_i+u_i) \, dV \nonumber \\ 
    &+ 
    \int_{CV} \frac{\partial}{\partial x_j} \rho (\overline{U}_i+u_i) (\overline{U}_j+u_j) \, dV
    \label{eq:reynolds_transport_momentum_decomped}
\end{align}
Applying the Reynolds averaging will result in the following equation:
\begin{align}
    \Sigma (\overline{F}_i) = \frac{\partial}{\partial t} \int_{CV} \rho (\overline{U}_i) \, dV 
    + 
    \int_{CV} \frac{\partial}{\partial x_j} \rho (\overline{U}_i\overline{U}_j+\overline {u_iu_j}) \, dV
    \label{eq:reynolds_transport_momentum_averaged}
\end{align}
The momentum equation along the $x_1$ direction, with the effective forces and with the simplifications related to the flow of interest, namely being fully developed, statistically stationary, incompressible, and two-dimensional, is as follows: 
\begin{equation}
    -\frac{\partial \overline{P}}{\partial x_1}Lhw-2\tau_{12w} Lw = \rho \int_{CV} \frac{\partial}{\partial x_2}  \overline {u_1u_2} \, dV
\end{equation}
where $\tau_{12w}=\mu \left.\frac{\partial \overline {U}_1}{\partial x_2}\right|_{w}$ is the viscous shear stress on the lower wall. Using the differential RANS equation, it can be shown that $\overline{u_1u_2}$ is negative at half the channel height, zero at the wall, and zero at the channel center. Due to the antisymmetry of the Reynolds stress profile about the channel center, the volume integral of the gradient of $\overline{u_1u_2}$ is zero.  Hence, knowing that the pressure gradient is constant in the flow direction, the pressure drop along the CV with length L reads:
\begin{eqnarray}
    \Delta P &=&  2\frac{L}{h}\underbrace{\mu \left. \frac{\partial \overline {U}_1}{\partial x_2}\right|_{w}}_{\tau_{12w}}
    \label{pressure_drop_from_momentum}
\end{eqnarray}
This equation clearly shows that the drop in pressure potential in the flow direction is used to overcome the viscous stresses on the walls. Nevertheless, the effect of turbulence is hidden in the velocity gradient of the mean longitudinal velocity. 

The impact of turbulence can be explicitly seen in the transport equation for the mechanical energy, which is 
\begin{eqnarray}
    \frac{D 1/2\, U_iU_i }{Dt}&=& U_i g_i-\frac{1}{\rho}U_i\frac{\partial P}{\partial x_i}+\frac{1}{\rho}U_i\frac{\partial \tau_{ij}}{\partial x_j}
    \label{eq:kinetic_energy1}
\end{eqnarray}
All dependent variables in this equation represent instantaneous values. The work done by the viscous forces per unit time (last term) can be rewritten as follows, so that the dissipation rate of mechanical energy becomes visible:
\begin{eqnarray}
    \frac{D 1/2\, U_iU_i }{Dt}&=& U_i g_i-\frac{1}{\rho}U_i\frac{\partial P}{\partial x_i}+
    \frac{1}{\rho}\frac{\partial U_i\tau_{ij}}{\partial x_j}-\frac{1}{\rho}\tau_{ij}\frac{\partial U_i}{\partial x_j}
    \label{eq:kinetic_energy2}
\end{eqnarray}
$\frac{1}{\rho}\tau_{ij}\frac{\partial U_i}{\partial x_j}$ is the dissipation rate, $\tau_{ij}=\mu(\partial U_i / \partial x_j+\partial U_j / \partial x_i)$ is the viscous stress tensor and $g_i$ is the gravitational acceleration. 

For Reynolds averaging, all dependent variables can be decomposed into ensemble mean and fluctuating components as follows:
\begin{eqnarray}
U_i&=&\overline{U_i}+u_i\\ \nonumber
P&=&\overline{P}+p\\ \nonumber
\tau_{ij}&=&\overline{\tau_{ij}}+\tau_{ij}^\prime
\end{eqnarray}
The overline symbol represents the ensemble averaging. Introducing decomposed variables into Eq.~\ref{eq:kinetic_energy2} and applying Reynolds Averaging, the ensemble mean of the kinetic energy equation reads:
\begin{align}
    &\frac{D 1/2\, \overline{U_i}\,\overline{U_i} }{Dt}
    +\frac{D 1/2\, \overline{u_iu_i}}{Dt} \nonumber \\ 
    &+\overline{u_iu_j}\frac{\partial \overline{U_i}}{\partial x_j}
    + \overline{U_i}\frac{\partial \overline{u_iu_j}}{\partial x_j}
    +\frac{1}{2}\frac{\partial \overline{u_iu_iu_j}}{\partial x_j} \nonumber \\
    &=\overline{U_i}g_i-\frac{1}{\rho}\overline{U_i}\frac{\partial \overline{P}}{\partial x_i}
    -\frac{1}{\rho}\overline{u_i\frac{\partial p}{\partial x_i}} \nonumber \\ 
    &+\frac{1}{\rho}\frac{\partial \overline{U_i}\overline{\tau_{ij}}}{\partial x_j}
    +\frac{1}{\rho}\frac{\partial \overline{u_i \tau_{ij}^\prime}}{\partial x_j}
    -\frac{1}{\rho}\overline{\tau_{ij}}\frac{\partial \overline{U_i}}{\partial x_j}
    -\frac{1}{\rho}\overline{\tau_{ij}^\prime\frac{\partial u_i}{\partial x_j}}
    \label{eq:RANS-kinetic_energy}
\end{align}
Note that $1/2\, \overline{U_i}\,\overline{U_i}$ and $1/2\, \overline{u_iu_i}$ are the kinetic energy per unit mass of the mean flow and the turbulent flow, respectively. This equation can be split into a mean and a turbulent component. The transport equation for the mechanical energy of the mean flow is:
\begin{align}
    &\frac{D 1/2\, \overline{U_i}\,\overline{U_i} }{Dt}
    + \overline{U_i}\frac{\partial \overline{u_iu_j}}{\partial x_j} \nonumber \\
    &=\overline{U_i}g_i-\frac{1}{\rho}\overline{U_i}\frac{\partial \overline{P}}{\partial x_i}
    +\frac{1}{\rho}\frac{\partial \overline{U_i}\overline{\tau_{ij}}}{\partial x_j}
    -\frac{1}{\rho}\overline{\tau_{ij}}\frac{\partial \overline{U_i}}{\partial x_j}
\label{eq:RANS-kinetic_energy_mean}
\end{align}
The second term on the left-hand side (LHS) is the extracted energy from the mean by turbulence, and the last term is the dissipation rate of the mean mechanical energy. The transport equation for the turbulent mechanical energy is:
\begin{align}
    &\frac{D 1/2\, \overline{u_iu_i}}{Dt} 
    +\overline{u_iu_j}\frac{\partial \overline{U_i}}{\partial x_j}
    +\frac{1}{2}\frac{\partial \overline{u_iu_iu_j}}{\partial x_j}\nonumber \\
    &=-\frac{1}{\rho}\overline{u_i\frac{\partial p}{\partial x_i}}
    +\frac{1}{\rho}\frac{\partial \overline{u_i \tau_{ij}^\prime}}{\partial x_j}
    -\frac{1}{\rho}\overline{\tau_{ij}^\prime\frac{\partial u_i}{\partial x_j}}
    \label{eq:RANS-kinetic_energy_turbulent}
\end{align}
The second and third terms at the LHS are the production of turbulence kinetic energy and the work done by the Reynolds stresses due to turbulent fluctuations, respectively.

The transport equation of mean mechanical energy (\ref{eq:kinetic_energy2}) for the two-dimensional horizontal fully developed incompressible statistically stationary turbulent channel flow simplifies to:
\begin{align}
    &\overline{u_iu_j}\frac{\partial \overline{U_i}}{\partial x_j}
    +\overline{U_i}\frac{\partial \overline{u_iu_j}}{\partial x_j}
    +\frac{1}{2}\frac{\partial \overline{u_iu_iu_j}}{\partial x_j}\nonumber \\
    &=-\frac{1}{\rho}\overline{U_i}\frac{\partial \overline{P}}{\partial x_i}
    -\frac{1}{\rho}\overline{u_i\frac{\partial p}{\partial x_i}} \nonumber \\
    &+\frac{1}{\rho}\frac{\partial \overline{U_i}\overline{\tau_{ij}}}{\partial x_j}
    +\frac{1}{\rho}\frac{\partial \overline{u_i \tau_{ij}^\prime}}{\partial x_j}
    -\frac{1}{\rho}\overline{\tau_{ij}}\frac{\partial \overline{U_i}}{\partial x_j}
    -\frac{1}{\rho}\overline{\tau_{ij}^\prime\frac{\partial u_i}{\partial x_j}}
    \label{eq:RANS-kinetic_energy_channel_diff}
\end{align}
Further simplification can be done when the volume integral of both sides of this equation for the CV in Fig.\ref{fig:CV} is done. In this integral form, some of the volume integrals can be converted to surface integrals using the divergence theorem. The result of this integration is as follows:
\begin{align}
    &\int_{CV}\overline{u_iu_j}\frac{\partial \overline{U_i}}{\partial x_j}dV
    +\int_{CV}\overline{U_i}\frac{\partial \overline{u_iu_j}}{\partial x_j}dV
    +\int_{CS} 1/2\overline{u_iu_iu_j}n_jdA \nonumber \\
    &=-\frac{1}{\rho}\int_{CS} \overline{P}\,\overline{U_i}n_i dA
    -\frac{1}{\rho}\int_{CS}\overline{p u_i}n_i dA \nonumber \\ 
    &+\frac{1}{\rho}\int_{CS}\overline{U_i}\,\overline{\tau_{ij}}n_j dA
    +\frac{1}{\rho}\int_{CS}\overline{u_i\tau_{ij}^\prime}n_j dA \nonumber\\ 
    &-\frac{1}{\rho}\int_{CV}\overline{\tau_{ij}}\frac{\partial \overline{U_i}}{\partial x_j}dV
    -\frac{1}{\rho}\int_{CV}\overline{\tau_{ij}^\prime\frac{\partial u_i}{\partial x_j}}dV 
    \label{eq:RANS-kinetic_energy_channel_integral1}
\end{align}
In the derivation for the pressure-related terms, the following continuity equations for the mean and fluctuating velocities  are used:
\begin{eqnarray}
   \frac{\partial \overline{U_i}}{\partial x_i}&=&0\\
   \frac{\partial u_i}{\partial x_i}&=&0
\end{eqnarray}
Because of the no-slip condition and the fully developed state of the flow,  the last three surface integrals have to be zero. 
\begin{align}
    &\int_{CV}\overline{u_iu_j}\frac{\partial \overline{U_i}}{\partial x_j}dV
    +\int_{CV}\overline{U_i}\frac{\partial \overline{u_iu_j}}{\partial x_j}dV \nonumber \\ 
    &=-\frac{1}{\rho}\int_{CS} \overline{P}\,\overline{U_i}n_i dA
    -\frac{1}{\rho}\int_{CV}\overline{\tau_{ij}}\frac{\partial \overline{U_i}}{\partial x_j}dV
    -\frac{1}{\rho}\int_{CV}\overline{\tau_{ij}^\prime\frac{\partial u_i}{\partial x_j}}dV 
    \label{eq:RANS-kinetic_energy_channel_integral2}
\end{align}
The LHS of the above equation represents the sum  of turbulence kinetic energy production and the extracted kinetic energy from the mean flow to turbulence. Thus, this sum can be converted to a surface integral, which is also zero owing to fully developed two-dimensional flow:
\begin{equation}
   \int_{CV}\overline{u_iu_j}\frac{\partial \overline{U_i}}{\partial x_j}dV
    +\int_{CV}\overline{U_i}\frac{\partial \overline{u_iu_j}}{\partial x_j}dV
    =\int_{CS}\overline{U_i}\overline{u_iu_j} n_j dA=0
    \label{eq:production_extration} 
\end{equation}
Hence, the production of turbulence kinetic energy is negative of the extracted energy from the mean flow. Hence, Eq.~\ref{eq:RANS-kinetic_energy_channel_integral2} reduces to the balance between the work done by the pressure and the viscous dissipation rate calculated from mean velocity gradients and the gradients of velocity fluctuations, respectively. 
\begin{equation}
    \int_{CS} \overline{P}\,\overline{U_i}n_i dA=
    -\int_{CV}\overline{\tau_{ij}}\frac{\partial \overline{U_i}}{\partial x_j}dV
    -\int_{CV}\overline{\tau_{ij}^\prime\frac{\partial u_i}{\partial x_j}}dV
    \label{eq:RANS-kinetic_energy_channel_integral4}
\end{equation}
The second term at the right-hand side (RHS) is the dissipation rate of turbulence kinetic energy, i.e., the mass integral of the last term in Eq.~\ref{eq:RANS-kinetic_energy_turbulent}. Using the divergence theorem and continuity equation for the LHS, and the two-dimensional nature of the flow, this equation can further be simplified to:
\begin{equation}
    \int_{CV} \frac{\partial \overline{P}}{\partial x_1}\,\overline{U_1}dV=
    -\int_{CV}\overline{\tau_{12}}\frac{\partial \overline{U_1}}{\partial x_2}dV
    -\int_{CV}\overline{\tau_{ij}^\prime\frac{\partial u_i}{\partial x_j}}dV
    \label{eq:RANS-kinetic_energy_channel_integral5}
\end{equation}
Writing the viscous stress term in terms of mean and fluctuating velocities delivers the final form of the energy balance in integral form as follows:
\begin{align}
    \int_{CV} \frac{\partial \overline{P}}{\partial x_1}\,\overline{U_1}dV=
    &-\mu\int_{CV} \left(\frac{\partial \overline{U_1}}{\partial x_2}\right)^2dV \nonumber \\
    &-\mu\int_{CV} \left(\overline{\frac{\partial u_i}{\partial x_j}\frac{\partial u_i}{\partial x_j}}
    +\overline{\frac{\partial u_j}{\partial x_i}\frac{\partial u_i}{\partial x_j}}
    \right)dV
    \label{eq:RANS-kinetic_energy_channel_integral6}
\end{align}
The last term represents the total dissipation rate of turbulence kinetic energy in the CV, and no further simplification can be done for it. 
When Eq.~\ref{eq:RANS-kinetic_energy_channel_integral6} is carefully examined,  one can end up with the conclusion that the pressure gradient is solely dependent on viscous effects, which is opposite to the observations that the coefficients of pressure drop or drag become independent of viscous effects at high Reynolds numbers \citep{Colebrook1939, Nikuradse1950, Idelchik2005}. This is a paradox that cannot be explained by Eqn.~\ref{eq:RANS-kinetic_energy_channel_integral6}. However, the transport equation of turbulent kinetic energy (\ref{eq:RANS-kinetic_energy_turbulent}) can resolve it. The integral of this equation for the flow considered here simplifies to 
\begin{equation}
   \int_{CV}\overline{u_iu_j}\frac{\partial \overline{U_i}}{\partial x_j}dV
    = -\frac{1}{\rho}\int_{CV}\overline{\tau_{ij}^\prime\frac{\partial u_i}{\partial x_j}}dV
    \label{eq:RANS-kinetic_energy_turbulent_integral}
\end{equation}
This equality clearly demonstrates that the production rate of turbulence kinetic energy is precisely counterbalanced by its dissipation rate. Moreover, the causal relationship established in the transport equation asserts that the production rate of turbulence kinetic energy directly dictates its dissipation rate. Hence, the turbulence dissipation rate in Eq.~\ref{eq:RANS-kinetic_energy_channel_integral6} can be replaced by the turbulence kinetic energy production as follows:
\begin{equation}
    \int_{CV} \frac{\partial \overline{P}}{\partial x_1}\,\overline{U_1}dV=
    -\mu\int_{CV} \left(\frac{\partial \overline{U_1}}{\partial x_2}\right)^2dV 
    +\rho \int_{CV}\overline{u_1u_2}\frac{\partial \overline{U_1}}{\partial x_2}dV
    \label{eq:RANS-kinetic_energy_channel_integral7}
\end{equation}
It is known that $\partial \overline{P}/\partial x_1$ is a constant. Hence, the integral on the LHS becomes
\begin{equation}
    \frac{\partial \overline{P}}{\partial x_1} \dot{Q}L=
    -\mu\int_{CV} \left(\frac{\partial \overline{U_1}}{\partial x_2}\right)^2dV 
    +\rho \int_{CV}\overline{u_1u_2}\frac{\partial \overline{U_1}}{\partial x_2}dV
    \label{eq:RANS-kinetic_energy_channel_integral8}
\end{equation}
where $\dot{Q}$ is the volume flow rate.  
The resulting relation clearly decomposes viscous and inertial turbulent effects. It reflects how the work done by the pressure gradient is channelled into viscous dissipation and turbulence production rates. This work is transformed irreversibly into internal energy. Furthermore, the viscous and inertial effects contributing to the wall shear stress become visible when $\partial \overline{P}/\partial x_1$ is left alone in this equation and equated to the one obtained from the momentum equation  (\ref{pressure_drop_from_momentum}) as follows: 
\begin{equation}
\tau_{12w}=
    \frac{\mu h}{2 \dot{Q}L}\int_{CV} \left(\frac{\partial \overline{U_1}}{\partial x_2}\right)^2dV 
    -\frac{\rho h}{2 \dot{Q}L} \int_{CV}\overline{u_1u_2}\frac{\partial \overline{U_1}}{\partial x_2}dV
    \label{eq:RANS-wall_shear_stress}
\end{equation}
Now, the relation between the wall shear stress, viscous dissipation, and turbulence production becomes apparent.  Knowing that the pressure gradient is constant in the flow direction, the pressure drop (Eq.~\ref{eq:RANS-kinetic_energy_channel_integral8}) can be written as:
\begin{equation}
    \Delta P=\underbrace{
    \frac{\mu}{\dot{Q}}\int_{CV} \left(\frac{\partial \overline{U_1}}{\partial x_2}\right)^2dV}_{\Delta P_{visc}} 
    \underbrace{-\frac{\rho}{\dot{Q}} \int_{CV}\overline{u_1u_2}\frac{\partial \overline{U_1}}{\partial x_2}dV}_{\Delta P_{turb}}
    \label{eq:RANS-kinetic_energy_channel_integral9}
\end{equation}
This equation is parallel to Colebrook's \citep{Colebrook1939} and Haaland's \citep{Haaland1983} friction factor relations, which have separate components for viscous and inertial turbulent effects.

\subsection{Order of Magnitude Analysis (OMA)}
The equations \ref{pressure_drop_from_momentum} and \ref{eq:RANS-kinetic_energy_channel_integral9} can be used to infer the pressure drop's scaling with respect to the other parameters. The former states that the wall shear stress determines the pressure gradient, whereas the latter states that the pressure drop arises due to viscous dissipation of the mean velocity and the production of turbulent kinetic energy.



As the pressure drop equation derived from the energy equation (Eq.~\ref{eq:RANS-kinetic_energy_channel_integral9}) explicitly captures the effect of turbulence, it will be used for OMA. At this point, several physically reasonable assumptions have to be made on the representative values of the unknown terms in Eq.~\ref{eq:RANS-kinetic_energy_channel_integral9}.    The gradients in $x_1$ and $x_2$ directions scale with $1/L$ and $1/h$, respectively. $\overline{U}_1$ scales with the ensemble and surface average mean velocity $\overline{U}_m$, and the $\overline{u_1u_2}$ scales with the negative of the surface average turbulent kinetic energy, which is $k_m=0.5(\overline{u_iu_i})$. Hence, the order of magnitudes of each term can be written as follows:
\begin{eqnarray}
\label{eq:order_of_magnitude_terms11}
\Delta P_{visc}& \approx & \frac{\mu}{\overline{U}_m w h}
\int_{CV} \left(\frac{\partial \overline{U_1}}{\partial x_2}\right)_m^2 \, dV \\
\label{eq:order_of_magnitude_terms12}
\Delta P_{turb}& \approx & \frac{\rho}{\overline{U}_m w h}
\int_{CV} k_m \left(\frac{\partial \overline{U}_1}{\partial x_2}\right)_m   \, dV 
\end{eqnarray}
The mean of the absolute value of the mean velocity gradient should reflect the effect of roughness on the change of the velocity profile. Thus, it is approximated by:
\begin{equation}
\label{eq:velocity_gradient_scaling}
\left|\frac{\partial \overline{U}_1}{\partial x_2}\right|_m \approx \frac{\overline{U}_m}{h}\left[A+B\left(\frac{\varepsilon}{h}\right)^k\right]    
\end{equation}
where $A$, $B$ and $k$ are unknown coefficients.

Assuming that the roughness ($\varepsilon$) is the key influencing factor for the level of generated turbulence, the mean turbulence kinetic energy ($k_m$) and the roughness can be related to each other by a kind of turbulence intensity as follows:
\begin{equation}
\label{eq:order_of_magnitude_kinetic_energy}
\frac{k_m}{\overline{U}_m^2} \approx \left[C+D\left(\frac{\varepsilon}{h}\right)^n\right]
\end{equation}
Here, a power-law relation is taken with an unknown power $n$, and $C$ and $D$ are unknown coefficients.

For the same type of flow in a pipe, $h$ can be replaced by the diameter $D$ of the pipe in the above relations. Hence, the orders of magnitude of the pressure drop terms in Eqs.~\ref{eq:order_of_magnitude_terms11} and ~\ref{eq:order_of_magnitude_terms12} are:
\begin{eqnarray}
\label{eq:order_of_magnitude_terms13}
\Delta P_{visc}& \approx & A_{visc}\frac{\mu \overline{U}_m}{D}
\frac{L}{D}\left[A+B\left(\frac{\varepsilon}{D}\right)^k\right]^2
\\
\label{eq:order_of_magnitude_terms14}
\Delta P_{turb}& \approx & A_{turb}\rho {\overline{U}_m^2} \frac{ L}{D}
\left[A+B\left(\frac{\varepsilon}{D}\right)^k\right]
\nonumber\\
& & \qquad\times \left[C+D\left(\frac{\varepsilon}{D}\right)^n\right]
\end{eqnarray}

where $A_{visc}$ and $A_{turb}$ are the correlation constants.

%
Normalizing this relation with $0.5\rho \overline{U}_m^2 L/D$ results in the following form of friction factor:
\begin{eqnarray}
\label{eq:order_of_magnitude_pressure_pipe_friction_factor1_visc}
f_{visc}& \approx & A_{visc}2\frac{1}{Re}
\left[A+B\left(\frac{\varepsilon}{D}\right)^k\right]^2
\\
\label{eq:order_of_magnitude_pressure_pipe_friction_factor1_turb}
f_{turb}& \approx & A_{turb}2  \left[A+B\left(\frac{\varepsilon}{D}\right)^k\right]\left[C+D\left(\frac{\varepsilon}{D}\right)^n\right]
\end{eqnarray}
A first approximation of total pressure drop and corresponding friction factor is the sum of Eqs.~\ref{eq:order_of_magnitude_terms13} and \ref{eq:order_of_magnitude_terms14}. Hence
\begin{align}
\label{eq:order_of_magnitude_pressure_pipe_final2}
\Delta P \approx &A_{visc} \mu \frac{L}{D}
\frac{\overline{U}_m}{D}\left[A+B\left(\frac{\varepsilon}{D}\right)^k\right]^2 \nonumber \\ 
&+ A_{turb}\rho \frac{L}{D}\overline{U}_m^2 \left[C+D\left(\frac{\varepsilon}{D}\right)^n\right]\left[A+B\left(\frac{\varepsilon}{D}\right)^k\right]
\end{align}
\begin{align}
\label{eq:order_of_magnitude_pressure_pipe_friction_factor2}
f \approx &A_{visc}2 \frac{1}{Re}\left[A+B\left(\frac{\varepsilon}{D}\right)^k\right]^2\nonumber \\ 
 &+ A_{turb} 2 \left[A+B\left(\frac{\varepsilon}{D}\right)^k\right] \left[C+D\left(\frac{\varepsilon}{D}\right)^n\right]
\end{align}
Note that $f_{visc}$ decreases with the Reynolds number and $f_{turb}$ changes only with the roughness as in Haaland's friction factor relation given below:
\begin{equation}
    \frac{1}{f^{1/2}}\approx-1.8 \log\left[ \frac{6.9}{Re}+\left(\frac{\varepsilon/D}{3.7}\right)^{1.11}\right]
\end{equation}

The following relations of \cite{Blasius1913} for  $\Delta P$ and $f$ hold for a smooth pipe at Reynolds numbers lower than $10^5$. 
\begin{eqnarray}
\label{eq:Blasius_DP}
\Delta P&\approx &0.158 L \rho^{3/4} \mu^{3/4} D^{-5/4} \overline{U}_m^{7/4}\\
\label{eq:Blasius_friction_factor}
f&\approx&0.316 Re^{-1/4}
\end{eqnarray}
Although this $\Delta P$ model lacks separate terms for the viscous and turbulent inertial terms, in principle, the power exponent of $7/4$ for $\overline{U}_m$ is expected since one can approximate a second-order polynomial, like the one in Eq.~\ref{eq:order_of_magnitude_pressure_pipe_final2}, by a power function having an exponent between 1 and 2.

The functional forms for the pressure drop and the friction factor (Eqs.~\ref{eq:order_of_magnitude_pressure_pipe_final2} and \ref{eq:order_of_magnitude_pressure_pipe_friction_factor2}) are derived by OMA, but they might not truly represent the exact behaviour. For example, the turbulent part takes a constant value at all Reynolds numbers. However, $f$ becomes Reynolds-number-independent only at large Reynolds numbers. In connection with that, the functional form of $f$ (Eq.~\ref{eq:order_of_magnitude_pressure_pipe_friction_factor2}) is not capable of reproducing the minima for rough pipes, which happens as $f$ transits from the smooth pipe regime to the rough pipe regime. This transition is also Reynolds-number-dependent.



It is important to note that taking the velocity gradient directly proportional to  $\overline{U}_m/h$ as shown in Eq.~\ref{eq:velocity_gradient_scaling} is an assumption. On the contrary, the $\overline{U}_m$ and $\mu$  in Eq.~\ref{eq:order_of_magnitude_terms11}, as well as   $\rho$  in Eq.~\ref{eq:order_of_magnitude_terms12} are not assumed, i.e., they are part of the derived governing equation. Hence, the power exponents of $\overline{U}_m$ in Eq.~\ref{eq:order_of_magnitude_pressure_pipe_final2} can be relaxed, only if the dimensional homogeneity requirement for $\Delta P$ model holds. The simplest way to guarantee the dimensional homogeneity is to make a symbolic regression to $f(Re, \varepsilon/D)$ instead of  $\Delta P (\overline U_m, \rho, \mu, L, D, \varepsilon)$.  
To have a transition from the smooth pipe regime to a rough pipe regime, the terms in the $f$ equation (\ref{eq:order_of_magnitude_pressure_pipe_friction_factor2}) should be relaxed to reflect the Reynolds number and relative roughness dependencies of this transition.  Allowing a transition behavior is legitimate because the assumption that dissipation is determined by production (Eq.~\ref{eq:RANS-kinetic_energy_turbulent_integral}) is expected to hold for rough pipe flows at high Reynolds numbers. 
The logistic function has previously been used and can reflect the transition \citep{Yang2009}. One form of the logistic function is:

\begin{equation}
\label{eq:logistic function}
g(t) = g_{lower}+\frac{g_{upper}-g_{lower}}{1+e^{-\beta(t-t_0)}}
\end{equation}
where $t$ is the independent variable, $\beta$ is the slope and $t_0$ is the midpoint of transition. Letting $g$ be $f$, $t$ be $Re$, and the upper and lower values to be $f_{turb}$ and $f_{visc}$ of Eqs.~\ref{eq:order_of_magnitude_pressure_pipe_friction_factor1_turb} and \ref{eq:order_of_magnitude_pressure_pipe_friction_factor1_visc}, respectively, a more general form of the friction factor can be obtained as follows:
\begin{align}
\label{eq:order_of_magnitude_pressure_pipe_friction_factor3}
f \approx &f_{visc}+\frac{f_{turb}-f_{visc}}{1+e^{-\beta(Re-Re_{tr})}} \nonumber \\
=&f_{visc}\frac{e^{-\beta(Re-Re_{tr})}}{1+e^{-\beta(Re-Re_{tr})}}+f_{turb}\frac{1}{1+e^{-\beta(Re-Re_{tr})}}
\end{align}
$Re_{tr}(\varepsilon/D)$ is the midpoint of transition and $\beta(\varepsilon/D)$ is the slope of the transition. These parameters are functions of relative roughness, and $\beta(\varepsilon/D)$ is expected to rise with increasing $\varepsilon/D$, whereas $Re_{tr}(\varepsilon/D)$ is expected to decrease. For the curve fitting, $Re_{tr}(\varepsilon/D)=Q/(\varepsilon/D)^r$
and $\beta=S(\varepsilon/D)^p$ can be used with the coefficients $Q$, $r$, $S$, and $p$. After relaxing the exponent of $Re$ from one to $\eta$, the functional form of the friction factor equation can be written as:
\begin{align}
\label{eq:order_of_magnitude_pressure_pipe_friction_factor4}
f \approx &\frac{2A_{visc}}{Re^\eta}
\left[A+B\left(\frac{\varepsilon}{D}\right)^k\right]^2
\frac{e^{-S (\varepsilon/D)^p[Re-Q/(\varepsilon/D)^r)]}}{1+e^{-S (\varepsilon/D)^p[Re-Q/(\varepsilon/D)^r)]}} \nonumber \\
&+A_{turb}2  \left[A+B\left(\frac{\varepsilon}{D}\right)^k\right]
\frac{\left[C+D\left(\frac{\varepsilon}{D}\right)^n\right]}{1+e^{-S (\varepsilon/D)^p[Re-Q/(\varepsilon/D)^r)]}}
\end{align}

\begin{table}
\caption{\label{table:OMA_coefficients}%
Curve-fit coefficients of Eq.~\ref{eq:order_of_magnitude_pressure_pipe_friction_factor4}.}
\begin{ruledtabular}
\begin{tabular}{l D{.}{.}{6} l D{.}{.}{6}}
Coefficient & \multicolumn{1}{c}{Value} &
Coefficient & \multicolumn{1}{c}{Value} \\
\hline
$A_{visc}$ & 0.8825   & $k$ & 0.4      \\
$A_{turb}$ & 0.6632   & $n$ & 0.5299   \\
$\eta$     & 0.2233   & $S$ & 0.003263 \\
$A$        & 0.5294   & $Q$ & 44.11    \\
$B$        & 0.006    & $p$ & 1.303    \\
$C$        & 0.03556  & $r$ & 0.844    \\
$D$        & 0.8237   & \multicolumn{1}{c}{--} & \multicolumn{1}{c}{--} \\
\end{tabular}
\end{ruledtabular}
\end{table}

\begin{figure}
  \centering
  \includegraphics[trim=3cm 0 1cm 0, clip, width=\columnwidth]{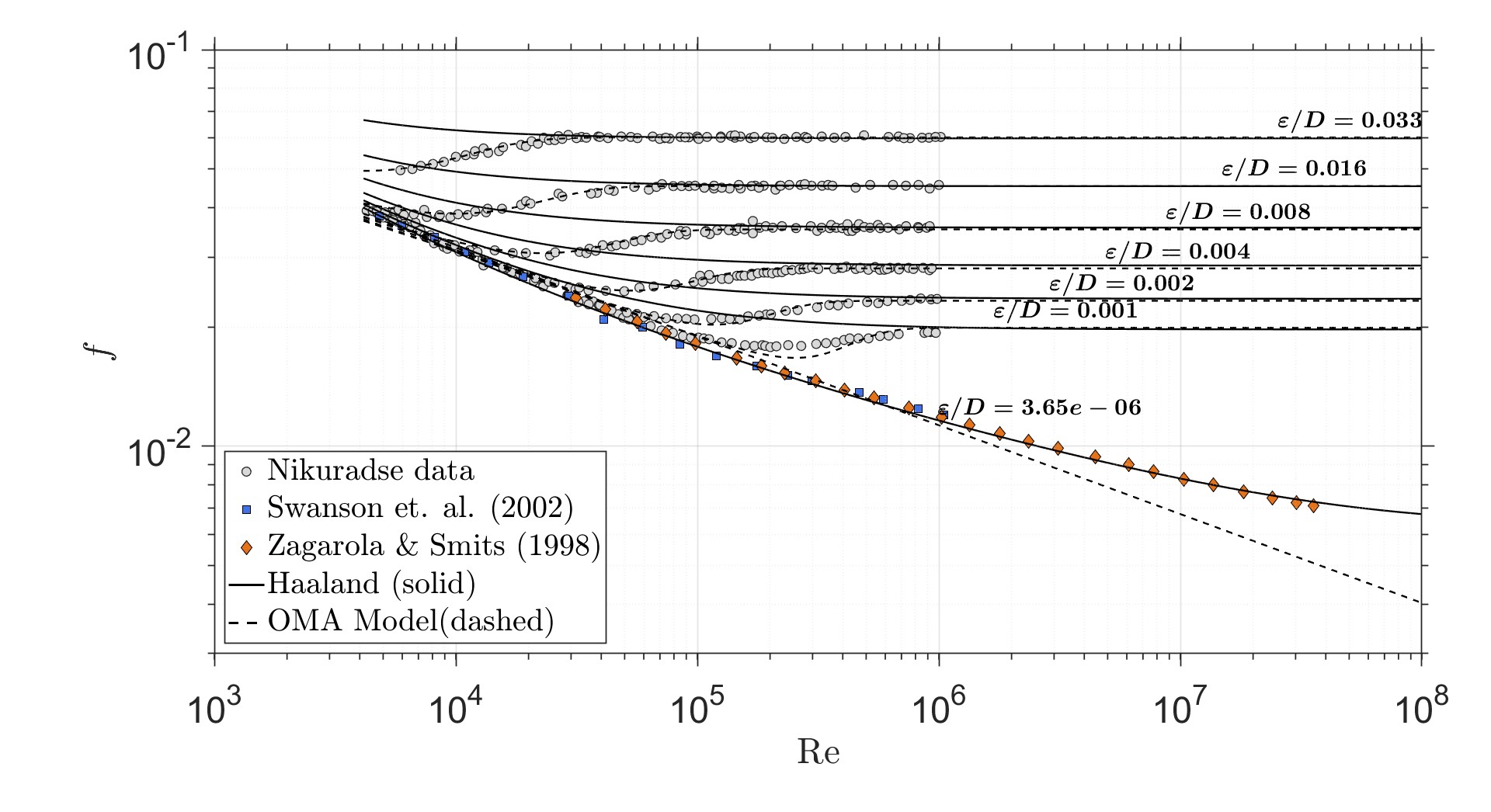}
  \caption{ OMA-based nonlinear regression to Eq.~\ref{eq:order_of_magnitude_pressure_pipe_final3}.}
  \label{fig:first_fit}
\end{figure}

This functional form with 13 coefficients is used to fit a nonlinear curve (Fig.~\ref{fig:first_fit}) to the low-Reynolds-number data of ~\cite{Nikuradse1950}, the high-Reynolds-number data of ~\cite{Swanson2002}, and the high-Reynolds-number data of ~\cite{Smits1998}. The constants of the curve fit are given in Table~\ref{table:OMA_coefficients}. The fit in Fig.~\ref{fig:first_fit} is very accurate up to $Re=10^6$, but it can not catch the high Reynolds number behavior of the smooth pipe as Colebrook's relation.


The resulting $\Delta P$ can be obtained simply multiplying the equation above by $L/D 0.5 \rho \overline{U}_m^2$ as follows:

\begin{align}
\label{eq:order_of_magnitude_pressure_pipe_final3}
\Delta P &= A_{visc}\,\rho^{1-\eta}\mu^{\eta}\,\frac{L}{D}\,
\frac{\overline{U}_m^{2-\eta}}{D^\eta}
\left[A+B\left(\frac{\varepsilon}{D}\right)^k\right]^2
\nonumber\\
&\quad\times
\frac{e^{-\beta(Re-Re_{tr})}}{1+e^{-\beta(Re-Re_{tr})}}
\nonumber\\
&\quad+ A_{turb}\,\rho\,\frac{L}{D}\,\overline{U}_m^2\,
\left[A+B\left(\frac{\varepsilon}{D}\right)^k\right]\,
\frac{\left[C+D\left(\frac{\varepsilon}{D}\right)^n\right]}{1+e^{-\beta(Re-Re_{tr})}}
\end{align}

It is not surprising to see that for $\eta=0.2233$, the exponents of OMA-based equations converge to those of Blasius in Eqs.~\ref{eq:Blasius_DP} and \ref{eq:Blasius_friction_factor} for smooth pipes.

\subsection{Constraints for Regression}\label{subsec:Constraints_for_Regression}
Despite their complexity and the extensive number of unknown parameters in Eqs.~
\ref{eq:order_of_magnitude_pressure_pipe_friction_factor3} and \ref{eq:order_of_magnitude_pressure_pipe_final3}, it is better to use these equations to guide symbolic regression alongside the experimental data. To bias the search toward physically admissible regressions for $f$, we incorporate four constraints ($C_1$-$C_4$), which are implemented in the regression process by using lightweight and model-only diagnostics to verify their fulfilment. These diagnostics are evaluated on synthetic grids of the independent variables. Using the friction factor regression models, the pressure drop is calculated over the synthetic grid, and the constraints are checked for fulfilment. The constraints are derived as follows:

\subsubsection{Constraint C\textsubscript{1}: Velocity Sensitivity Exponent}

According to Eqs.~\ref{eq:order_of_magnitude_pressure_pipe_friction_factor4}
and \ref{eq:order_of_magnitude_pressure_pipe_final3},
the dependence of $\Delta P$ on the mean speed \(\overline U_m\) at fixed
\(\rho,\mu,\varepsilon,D\), and \(L\) is a linear–plus–quadratic superposition.
At sufficiently small  \(\overline U_m\) the viscous term dominates and
\(\Delta P\propto \overline U_m\); at sufficiently large \(\overline U_m\)
the turbulent term dominates and \(\Delta P\propto \overline U_m^{2}\).
We cannot control the exact shape of the SR model during regression, but
locally, it can be approximated by a power law
\begin{equation}
\label{eq:DP-Um_model}
    \Delta P_{\text{model}} \approx R \,\overline{U}_m^\chi .
\end{equation}
A local logarithmic velocity exponent \(\chi\) is then defined as
\begin{equation}
\label{eq:nUdef}
\chi=\frac{\mathrm \partial \log \Delta P_{\text{model}}}{\mathrm \partial\log \overline{U}_m}
\;=\;\frac{\overline{U}_m}{\Delta P_{\text{model}}}\,
      \frac{\partial \Delta P_{\text{model}}}{\partial \overline{U}_m} .
\end{equation}
Evaluating \(\chi\) for the OMA model over the Nikuradse
data range shows that it smoothly interpolates between
\(\chi\approx1\) in the viscous limit and values slightly above
\(\chi=2\) in the smooth–rough transition, before relaxing back toward
\(\chi=2\) in the fully rough regime.  In particular, the OMA model
remains within
\(1 \lesssim \chi \lesssim \chi_{\max}\) with
\(\chi_{\max}\simeq 2.4\).
We therefore use the following properties as constraint $C_1$:
\begin{subequations}
\label{eq:C1-props}
\begin{equation}
\Delta P_{\text{model}} \ \text{is non-decreasing in } \overline U_m,
\end{equation}
\begin{equation}
1 \;\lesssim\; \chi \;\lesssim\; \chi_{\max},
\end{equation}
\begin{equation}
\lim_{\overline U_m\to\infty} \chi = 2 .
\end{equation}
\end{subequations}
Here \(\chi_{\max}\) is taken directly from the OMA prediction; it is
used as a soft upper bound during regression rather than a hard
universal limit.

\begin{figure}
  \centering
  \includegraphics[width=\columnwidth]{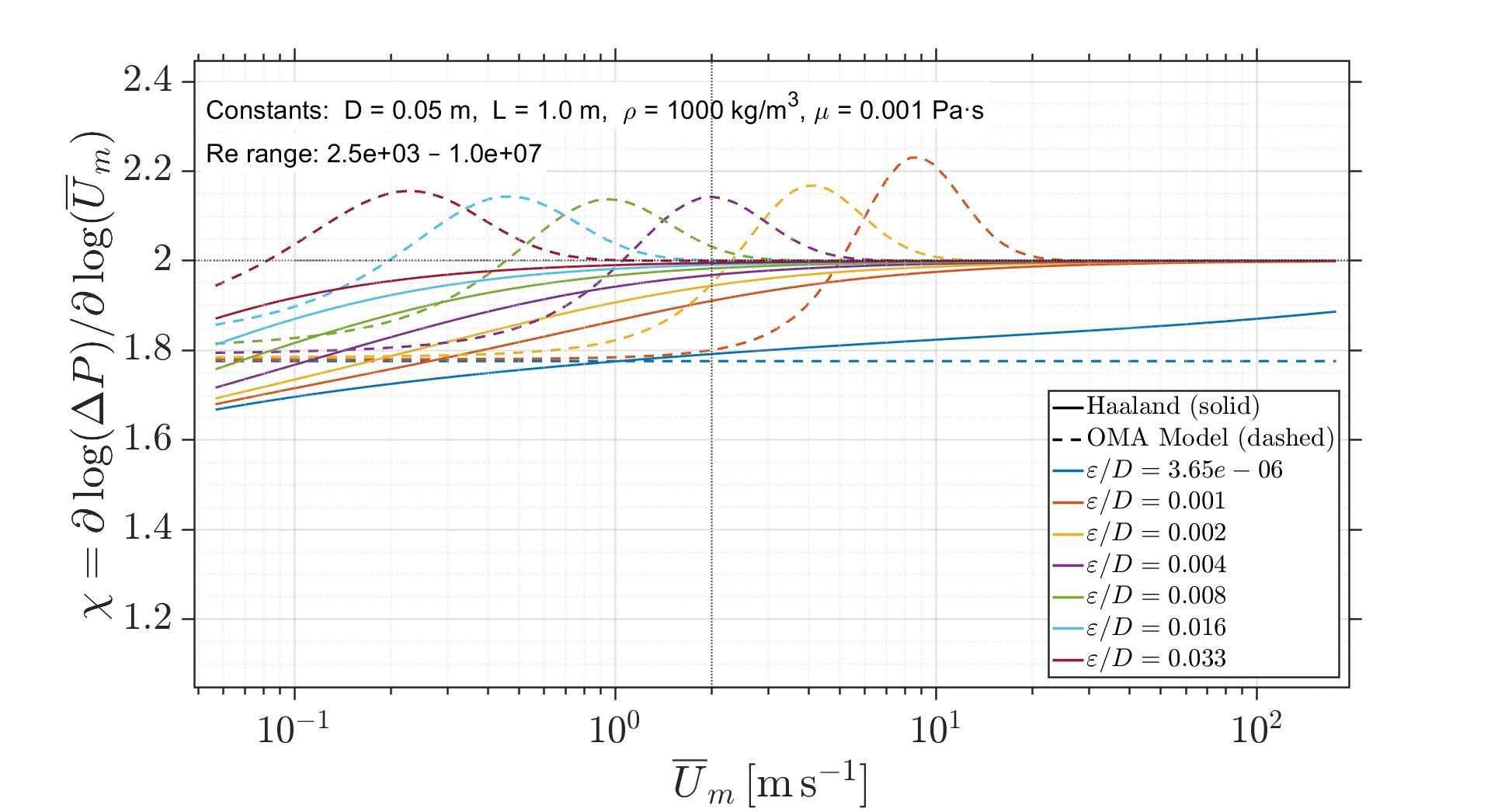}
  \caption{\label{fig:OMA_C1}Exponent of surface average mean velocity of OMA-based model and Haaland's model.}
\end{figure}

\subsubsection{Constraint C\textsubscript{2}: Sensitivity to Roughness at Fixed Reynolds Number}

In the OMA model (Eq. \ref{eq:order_of_magnitude_pressure_pipe_friction_factor4}), wall roughness enters only through
increasing factors of the form \((\varepsilon/D)^k\) and \((\varepsilon/D)^n\) with
\(k,n\ge 0\). At fixed \(\overline{U}_m,\rho,\mu,D,L\) this implies that the
pressure drop must not \emph{systematically} decrease when the pipe is made rougher:
Once the flow is in the hydraulically rough regime, additional roughness can only
increase the resistance.
We quantify the roughness sensitivity through the logarithmic exponent:
\begin{equation}
  s \;=\; \frac{\mathrm \partial \log \Delta P}{\mathrm \partial \log \varepsilon}
  \;=\; \frac{\varepsilon}{\Delta P}\,
        \frac{\partial \Delta P}{\partial \varepsilon}.
\end{equation}
When evaluated on the OMA model over Nikuradse-like conditions, $s$ is
non–negative and remains below a modest upper bound
$s_{\max}\approx 0.4$. This motivates the following
testable properties at fixed Reynolds number:
\begin{subequations}
\label{eq:C2-props}
\begin{equation}
\frac{\partial \Delta P}{\partial \varepsilon} \;\ge\; 0,
\end{equation}
\begin{equation}
0 \;\le\; s
      \;=\; \frac{\mathrm \partial \log \Delta P}{\mathrm \partial \log \varepsilon}
      \;\lesssim\; s_{\max} \approx 0.4 .
\end{equation}
\end{subequations}
Small, localized negative values of \(s\) near the hydraulically smooth
limit are interpreted as numerical artifacts of the finite-difference
evaluation and are tolerated to a limited extent. The precise aggregation
of violations into a scalar $C_2$ score is described in the constraint–scoring
section.

\begin{figure}
  \centering
  \includegraphics[width=\columnwidth]{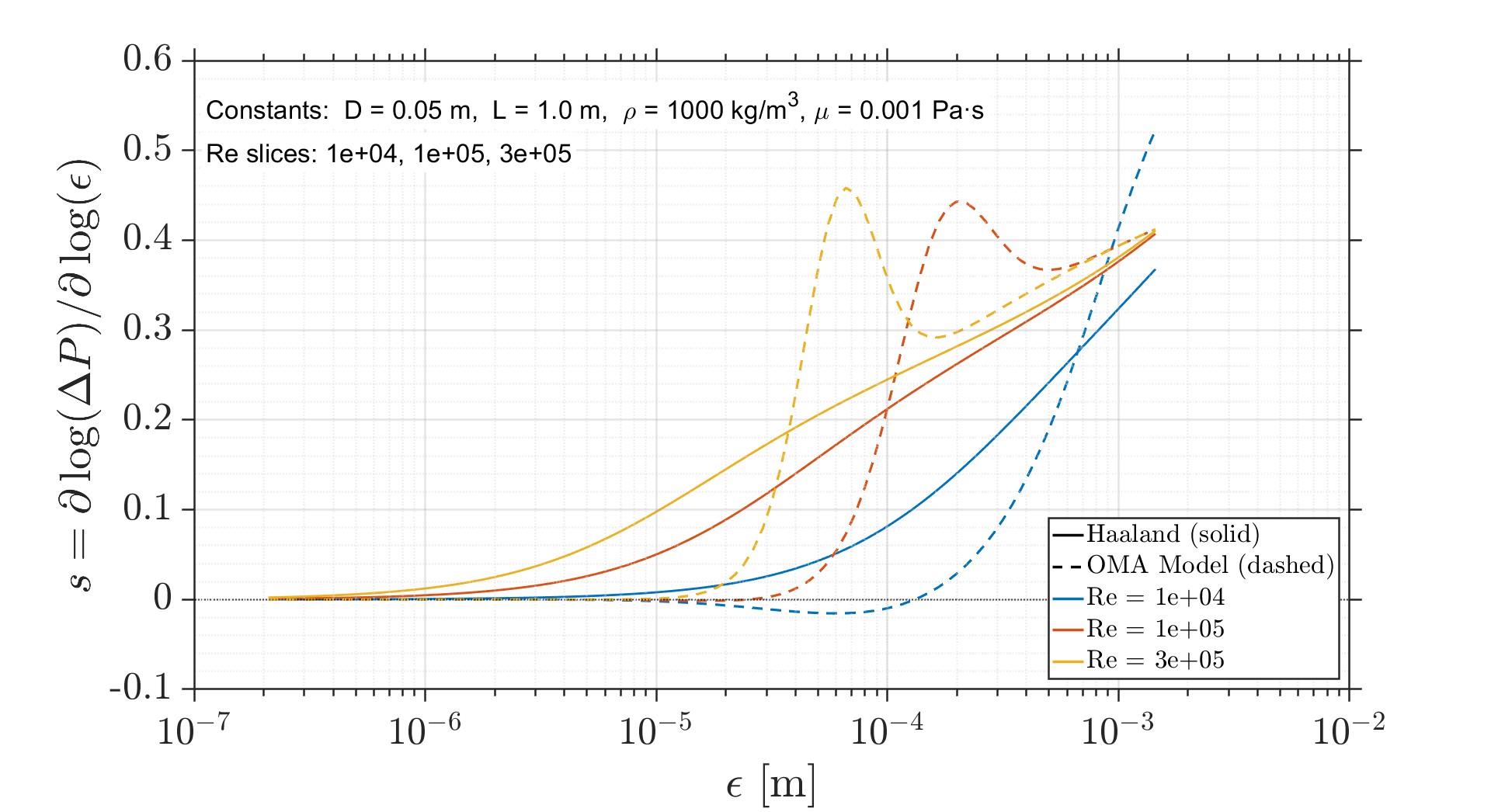}
  \caption{\label{fig:OMA_C2}Exponent of roughness of OMA-based model and Haaland's model.}
\end{figure}

\subsubsection{Constraint C\textsubscript{3}: Viscous Consistency at Fixed Velocity}

At fixed \(\overline{U}_m,\rho,D,L\) the OMA model contains a viscous
contribution that is linear in \(\mu\) and a turbulent contribution that
does not depend explicitly on \(\mu\).  In the hydraulically smooth
regime, this yields a small positive logarithmic exponent, whereas in the
fully rough regime, the pressure drop becomes effectively independent of
\(\mu\).
We measure this with:
\begin{equation}
\alpha \;=\; \frac{\mathrm \partial\log \Delta P}{\mathrm \partial\log \mu}
\;=\; \frac{\mu}{\Delta P}\,\frac{\partial \Delta P}{\partial \mu}.
\end{equation}
For the OMA model, \(\alpha\) lies in a narrow positive range
\(0 \lesssim \alpha \lesssim 0.3\) and approaches \(\alpha\to 0\) as
\(\varepsilon/D\to\infty\).
The SR equation preserves the same qualitative trend, but can show
mildly negative \(\alpha\) for very rough pipes and low Reynolds
numbers, where the explicit roughness dependence dominates and
increasing \(\mu\) shifts the operating point along a nearly flat
portion of the rough plateau.
Guided by the OMA envelope, we therefore adopt:
\begin{subequations}
\label{eq:C3-props}
\begin{equation}
\alpha_{\min} \;\le\; \alpha \;\le\; \alpha_{\max},
\end{equation}
\begin{equation}
\lim_{Re\to\infty} \alpha = 0,
\end{equation}
\end{subequations}
with \(\alpha_{\min}\approx -0.5\) and \(\alpha_{\max}\approx 1\).
These values are wide enough to contain the SR model wherever it
matches the OMA behaviour, while excluding unphysical strong decreases
or increases of \(\Delta P\) with viscosity.

\begin{figure}
  \centering
  \includegraphics[width=\columnwidth]{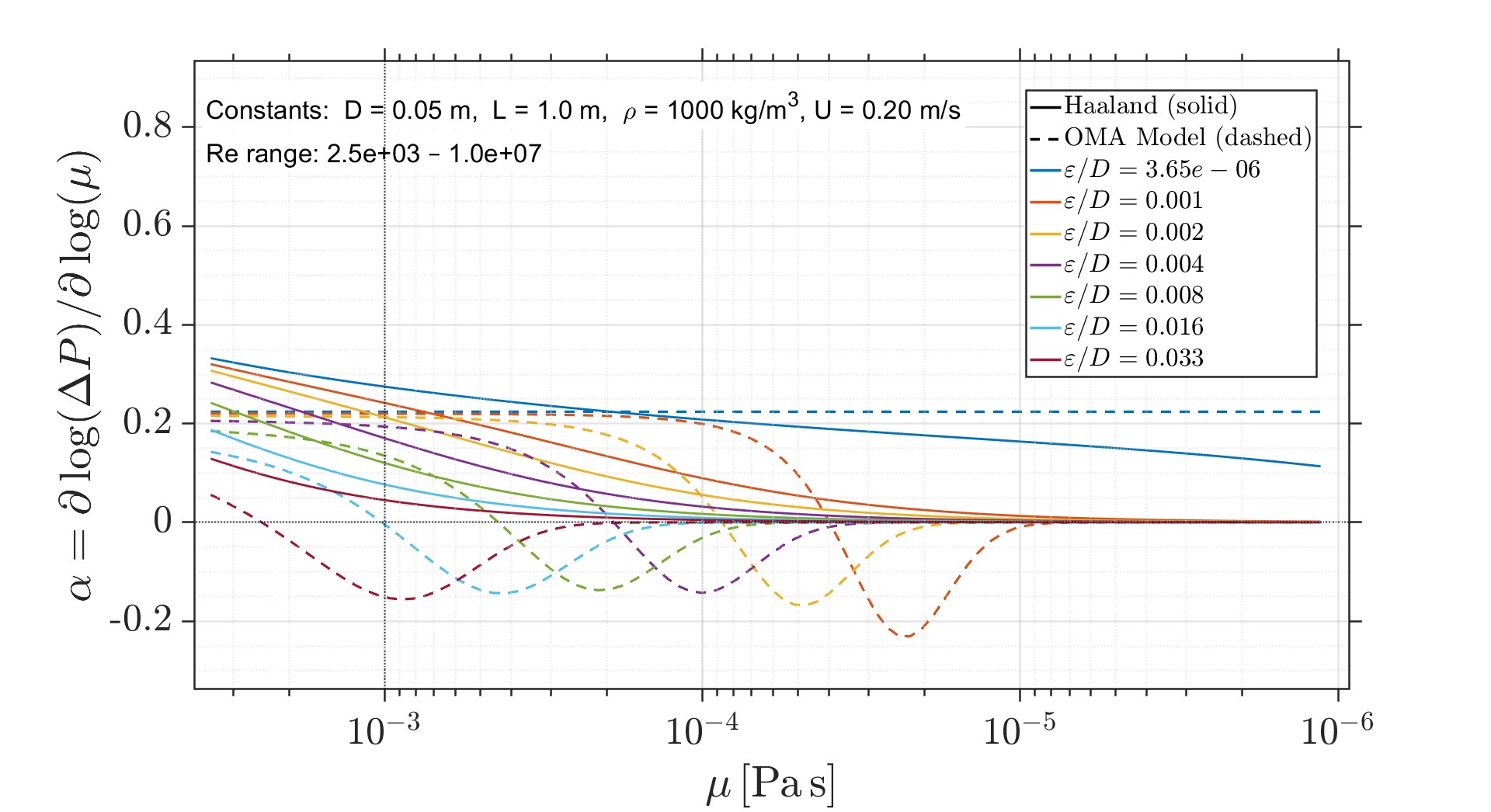}
  \caption{\label{fig:OMA_C3}Exponent of viscosity of OMA-based model and Haaland's model.}
\end{figure}

\subsubsection{Constraint C\textsubscript{4}: Density Sensitivity at Fixed Velocity}

In the OMA model, at fixed \(\overline{U}_m,\mu,D,L\) the turbulent
contribution is linear in \(\rho\) whereas the viscous term is
independent of \(\rho\).  Neglecting the weak density–dependence of the
friction factor, this suggests a logarithmic exponent of order unity:
\begin{equation}
\gamma \;=\; \frac{\mathrm \partial\log \Delta P}{\mathrm \partial\log \rho}
\;=\; \frac{\rho}{\Delta P}\,\frac{\partial \Delta P}{\partial \rho}.
\end{equation}
When evaluated on the OMA model over the turbulent range,
\(\gamma\) stays close to one, typically
\(0.7 \lesssim \gamma \lesssim 1\), and approaches \(\gamma\to1\) as the
flow becomes fully rough.
The SR correlation reproduces this structure but, in the smooth–rough
transition, the implicit \(\rho\)-dependence through \(\mathrm{Re}\) can
reinforce the explicit \(\rho\) factor and yield exponents up to
\(\gamma\approx 1.3\).  Using OMA as the physical reference, we thus
relax the upper bound while enforcing positivity of
\(\partial\Delta P/\partial\rho\):
\begin{subequations}
\label{eq:C4-props}
\begin{equation}
0 \;\le\; \gamma \;\le\; \gamma_{\max},
\end{equation}
\begin{equation}
\frac{\partial \Delta P}{\partial \rho} \;\ge\; 0,
\end{equation}
\end{subequations}
with \(\gamma_{\max}\approx 1.4\).
This keeps the effective density exponent in the OMA–consistent
order–unity range while preventing any unphysical decrease of pressure
drop with increasing density.

\begin{figure}
  \centering
  \includegraphics[width=\columnwidth]{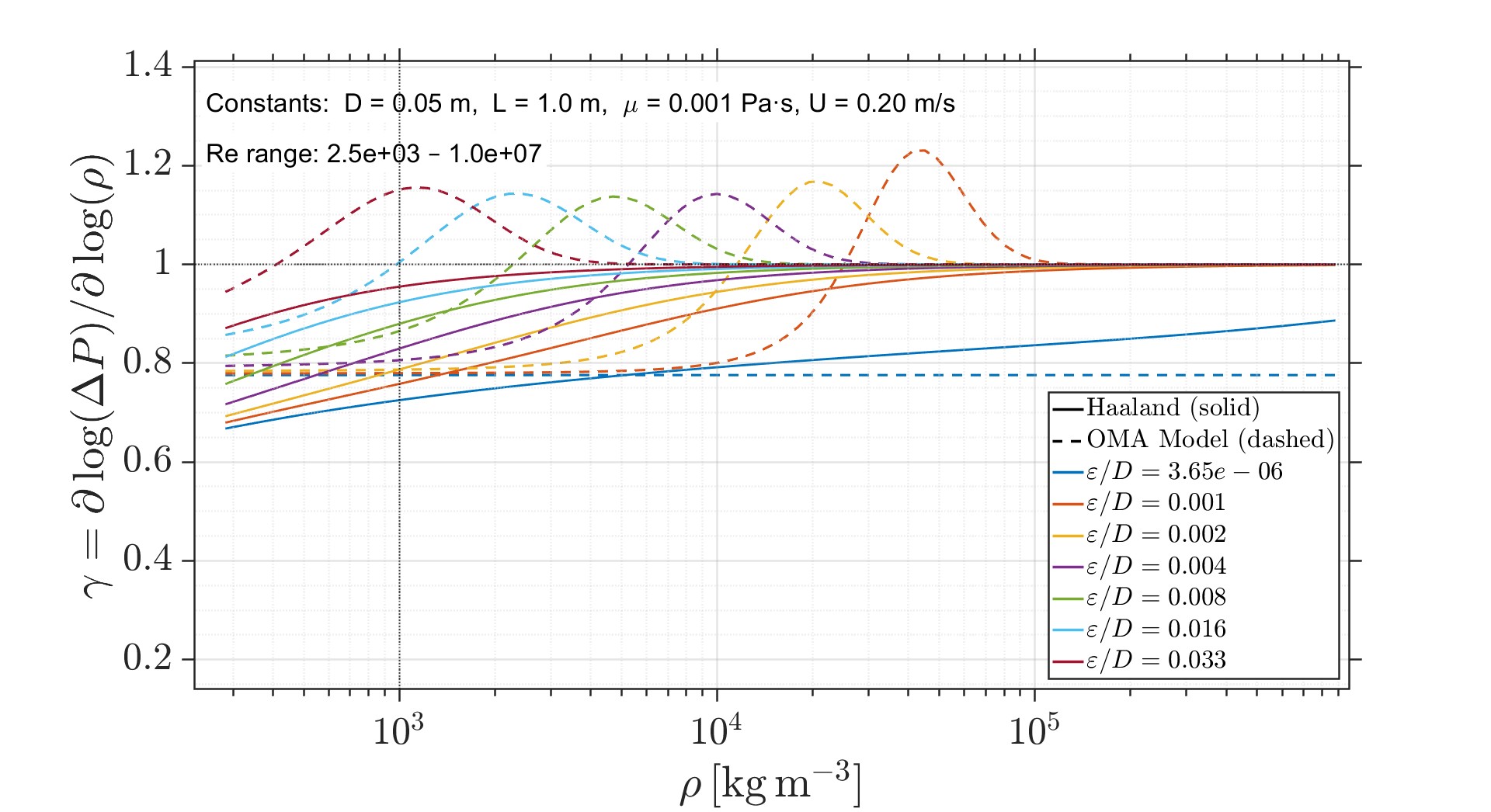}
  \caption{\label{fig:OMA_C4}Exponent of density of OMA-based model and Haaland's model.}
\end{figure}

\subsection{Symbolic Regression with Constraints}

We implement symbolic regression by modifying the multi-gene genetic programming
framework of GPTIPS 2.0 \citep{Searson2015}. The most obvious modifications are the conversion from 2-D to 3-D optimization and the new joint coefficient optimization method. Each individual model is a linear combination of $G$ expression trees (genes):
\[
  f(\mathrm{Re},\varepsilon/D)
  \;=\;
  \sum_{g=1}^{G} w_g\,T_g\big(\mathrm{Re},\varepsilon/D;\,\theta_g\big),
\]
where each gene $T_g$ is built from a primitive set
$\{+,-,\times,/,(\cdot)^m,\log,\exp,\tanh\}$, terminal nodes are
dimensionless inputs from the Buckingham--$\Pi$ grid, and $\theta_g$ collects
ephemeral (numeric) constants inside the trees.

\par\medskip\noindent\textbf{Population initialization.}\quad
The initial population is generated by ramped half–and–half with prescribed
depth and node limits. Domain–safe primitives (protected division, safe
logarithms, etc.) are used so that every expression can be evaluated on the
entire training grid $(\mathrm{Re},\varepsilon/D)$ without producing
non–finite values.

\par\medskip\noindent\textbf{Joint optimization of weights and constants.}\quad
For each individual, all numeric quantities that can be tuned are collected
into a single parameter vector
\[
  par \;=\; \big(w_1,\dots,w_G,\;\theta_1,\dots,\theta_G\big).
\]
Optimization proceeds in two stages. First, with constants $\theta_g$ fixed at
their current values, all genes are evaluated on the training set to form a
design matrix $G$ whose columns contain the gene outputs. A ridge regression on
$G$ provides an initial estimate of the weights $w_g$; any non-finite columns
are zeroed so that pathological genes are automatically assigned zero weight.
Second, the warm-started weights are concatenated with the current constants to
form an initial vector $par_0$, and a Nelder--Mead search is applied to minimize
the prediction error of $\Delta P$ on the training data. Unlike standard GPTIPS
2, this step jointly refines both linear coefficients and non-linear constants,
which helps prevent extremely nested functions while improving asymptotic
behaviour and agreement with Nikuradse trends.

\par\medskip\noindent\textbf{Three objective components.}\quad
In our modified version of GPTIPS 2.0, using the optimized parameter vector $par^\star$, each individual is evaluated in
terms of three minimization objectives:
\begin{itemize}
  \item \textbf{Fitness} $J_{\mathrm{err}}$: root–mean–squared error
        (RMSE) of the predicted $\Delta P$ on the training set.
 \item \textbf{Complexity} $J_{\mathrm{comp}}$: the total number of nodes
      (operators, variables, and constants) in the final symbolic expression.
      Since our models are multi-gene, we compute $J_{\mathrm{comp}}$ by
      summing the node counts of all gene trees. In this paper, this value is
      reported simply as ``nodes''.
  \item \textbf{Physics objective} $J_{\mathrm{phys}}$: we compute the four
        constraint scores $C_1$--$C_4$ from the order–of–magnitude analysis
        (velocity exponent, roughness monotonicity, viscous consistency,
        density sensitivity). Each $C_k\in[0,1]$ measures violation of the
        corresponding OMA trend, and we aggregate them as
        \[
          J_{\mathrm{phys}} \;=\; \max\{C_1,C_2,C_3,C_4\},
        \]
        so that a model is only considered strongly physics–consistent if it
        behaves correctly in all four directions.
\end{itemize}
This yields a three–dimensional objective vector
\[
  \boldsymbol{J}
  \;=\;
  \big(J_{\mathrm{fit}},\,J_{\mathrm{comp}},\,J_{\mathrm{phys}}\big)
\]

\par\medskip\noindent\textbf{Numerical evaluation of the physics scores.}\quad
For any candidate friction-factor model \(f_{\text{model}}(Re,\varepsilon/D)\),
we evaluate it as a pressure-drop predictor via
\[
\Delta P_{\text{model}}
= f_{\text{model}}(Re,\varepsilon/D)\;
\frac{\rho\,\overline{U}_m^{2}\,L}{2D},
\]
Constraint scores are computed from logarithmic derivatives estimated using
finite-difference or gradient operators on log-spaced grids; any non-finite or
non-positive model output encountered during a sweep is treated as a hard
failure and yields the maximal penalty for that sweep.

Score $C_1$ enforces velocity-scaling consistency. For each tested
$\varepsilon/D$, we sweep $Re\in[10^{4},10^{9}]$ on a log grid (160 points),
map to $\overline U_m = Re\,\mu_{\mathrm{ref}}/(\rho D)$, and estimate
\[
\chi=\frac{\partial\log \Delta P_{\text{model}}}{\partial\log \overline U_m}
\]
using \texttt{gradient} applied to $\log \Delta P$ versus $\log \overline U_m$
(after a light three-point moving-average smoothing). We apply a two-zone
envelope: in the transitional range $10^{4}\le Re\le 10^{6}$ we require
$1.0\le\chi\le2.4$, while in the fully rough tail $10^{6}\le Re\le 10^{9}$ we
enforce $\chi\approx2$ within tolerance $\texttt{tolB}=0.01$. The two penalties
are combined as $0.3\,\text{penA}+0.7\,\text{penB}$ per roughness level, and
$C_1$ is taken as the worst case over $\varepsilon/D$.

Score $C_2$ enforces roughness monotonicity. At 12 Reynolds-number slices
$Re\in[3\times10^{3},10^{8}]$, we sweep the physical roughness $\varepsilon$ on a
log grid (360 points) spanning:

$\varepsilon\in[0.5\,(\varepsilon/D)_{\min}D,\;1.5\,(\varepsilon/D)_{\max}D]$ and
compute
\[
s=\frac{\partial\log \Delta P_{\text{model}}}{\partial\log \varepsilon}
\]
using \texttt{gradient}. We tolerate small negative slopes via
$s\ge -s_{\text{allow}}$ with $s_{\text{allow}}=0.1$, and define $C_2$ by
combining the fraction of violating grid points across $\varepsilon$ and then
averaging across Reynolds slices.

Score $C_3$ enforces viscosity consistency. At fixed $\varepsilon/D=10^{-3}$ we
sweep $\mu\in[10^{-4},10^{-1}]$ (120 log-spaced points) while keeping
$\overline U_m=2\,\mathrm{m/s}$ by setting $Re=\rho\overline U_m D/\mu$. We then
compute $\Delta P_{\text{model}}(\mu)$ and set $C_3$ to the fraction of points
where $\Delta P$ decreases with increasing $\mu$ (zero tolerance in the current
implementation).

\[
\alpha=\frac{\partial\log \Delta P_{\text{model}}}{\partial\log \mu}
\]

Score $C_4$ enforces density consistency. At fixed $\varepsilon/D=10^{-3}$ and
$\overline U_m=2\,\mathrm{m/s}$ we sweep $\rho\in[10^{2},10^{4}]$ (20 log-spaced
points) and estimate
\[
\gamma=\frac{\partial\log \Delta P_{\text{model}}}{\partial\log \rho}.
\]
We take $C_4=\langle|\gamma-1|\rangle$, with invalid evaluations assigned $C_4=1$.
The overall physics objective is reported as $J_{\mathrm{phys}}=\max(C_1,C_2,C_3,C_4)$.

Finally, the overall physics objective reported in the Pareto analysis is
\[
J_{\mathrm{phys}}=\max(C_1,C_2,C_3,C_4),
\]
so that any strong violation in one channel dominates the physics score.

\par\medskip\noindent\textbf{3-D Pareto selection.}\quad
We use Pareto dominance and crowding distance in the three dimensional objective
space $(J_{\mathrm{err}},J_{\mathrm{comp}},J_{\mathrm{phys}})$: individuals are
ranked into Pareto fronts, crowding distance promotes coverage of the trade-off
surface, and tournament selection favours non-dominated individuals with large
crowding distance while maintaining predictive pressure toward low
$J_{\mathrm{err}}$. Crucially, $J_{\mathrm{phys}}$ is not folded into the fitness
as a weighted penalty; treating physics as an independent axis avoids arbitrary
weighting and yields cleaner trade-off.

\par\medskip\noindent\textbf{Variation, elitism and termination.}\quad
From the selected parents, crossover, mutation, and direct reproduction are applied with prescribed probabilities to form the next generation; a small elite subset of the current Pareto front is copied unchanged. After a fixed number of generations, the final 3-D Pareto front is exported, and candidate correlations are chosen by inspecting the trade–off between fitness, complexity, and physicality within the constraints.

\begin{figure*}
\centering
\begin{tikzpicture}[
  >=Stealth,
  every node/.style={font=\small},
  box/.style={
    draw,
    rounded corners,
    align=center,
    inner sep=4pt,
    outer sep=0pt
  }
]

\node[box] (init) {Initial population};
\node[box, below=6mm of init] (decode) {Decode individual\\tree $\rightarrow$ gene templates};
\node[box, below=6mm of decode] (constopt) {Constant \& weight optimisation\\
(ridge LS $\rightarrow$ initial $\theta$,\\
Nelder--Mead on $[\theta,C]$)};
\node[box, below=6mm of constopt] (pred) {Evaluate on fitness $J_{\text{err}}$};

\node[box, below=6mm of pred] (phys) {Physics checks\\
compute $C_1,\dots,C_4$ and\\
$J_{\text{phys}}=\max(C_1,\dots,C_4)$};

\node[box, right=32mm of phys, anchor=west, yshift=38mm] (obj) {Form 3D objective vector\\
$(J_{\text{err}},\,J_{\text{comp}},\,J_{\text{phys}})$};

\node[box, below=6mm of obj] (pareto) {Multi-objective selection\\
(Pareto sort + crowding)\\
+ variation (crossover, mutation,\\
direct reproduction)};
\node[box, below=6mm of pareto] (next) {New population};

\draw[->] (init) -- (decode);
\draw[->] (decode) -- (constopt);
\draw[->] (constopt) -- (pred);
\draw[->] (pred) -- (phys);

\path (phys.east) ++(12mm,0) coordinate (xMid);

\draw[->] (phys.east) -- (xMid) |- (obj.west);

\draw[->] (obj) -- (pareto);
\draw[->] (pareto) -- (next);

\path (pareto.east) ++(16mm,0) coordinate (xR);
\draw[->] (next.east) -- (xR |- next.east)
                  -- (xR |- decode.east)
                  -- (decode.east);

\end{tikzpicture}
\caption{Workflow of the symbolic regression with physics constraints: each individual
undergoes joint optimisation of gene weights and constants, physics checks
\(C_1\)–\(C_4\), and 3D Pareto selection on \((J_{\mathrm{err}},J_{\mathrm{comp}},J_{\mathrm{phys}})\)
over successive generations.}
\end{figure*}
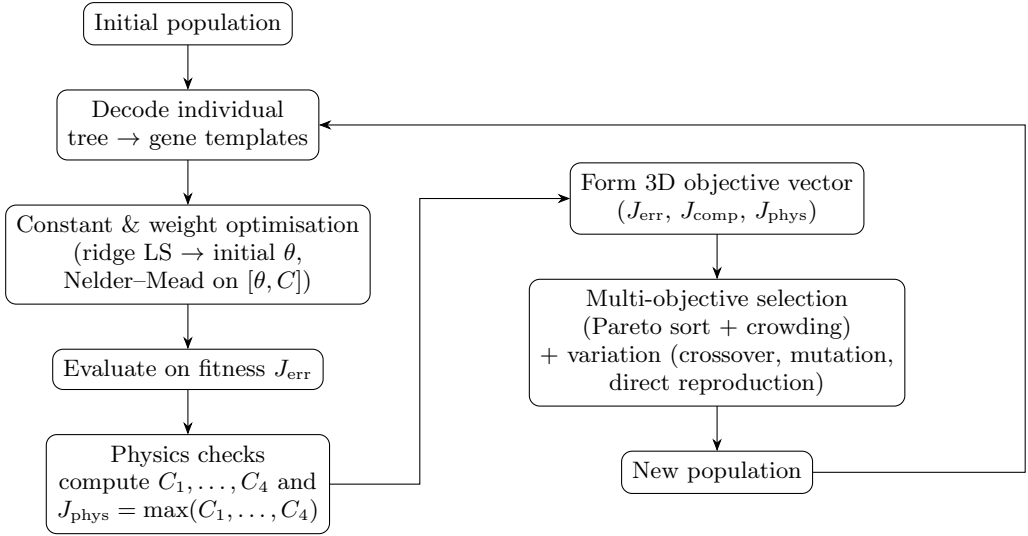

\section{Results}
\subsection{Pareto Trade-off}

The multi-objective search produces a final population evaluated by three criteria:
the data fitness \(J_{\mathrm{err}}\), the structural complexity \(J_{\mathrm{comp}}\) (node count),
and the physics score 
\\
\(J_{\mathrm{phys}}=\max(C_1,\ldots,C_4)\), where smaller values are preferable in all objectives.
Rather than visualizing the outcome in a three dimensional scatter space, we report the results as objective values versus population index and as robustness summaries for selected candidates.
Each point corresponds to one symbolic individual after convergence of the multi-objective GP run; lower values are preferable in all three coordinates. Blue markers denote all individuals, while red markers indicate the non-dominated subset (Pareto front).
This representation makes it easier to identify clusters of high-performing solutions and to locate specific candidates by index, while revealing the expected trade-off: individuals with very low \(J_{\mathrm{err}}\) often require higher complexity \(J_{\mathrm{comp}}\) or exhibit weaker physics compliance (larger \(J_{\mathrm{phys}}\)),
whereas strongly physics-compatible solutions typically sacrifice some predictive accuracy.
From the Pareto set we selected four representative candidates (Table~\ref{tab:candidates}),
spanning different compromises among error, complexity, and physics score. Two different runs, namely run A and run B, are examined to extract candidate equations. Run B has a stricter constraint check procedure compared to Run A.

\begin{figure*}[!hbt]
  \centerline{\includegraphics[
    trim={1cm 0 1cm 0},clip,
    width=0.7\textwidth,
  ]{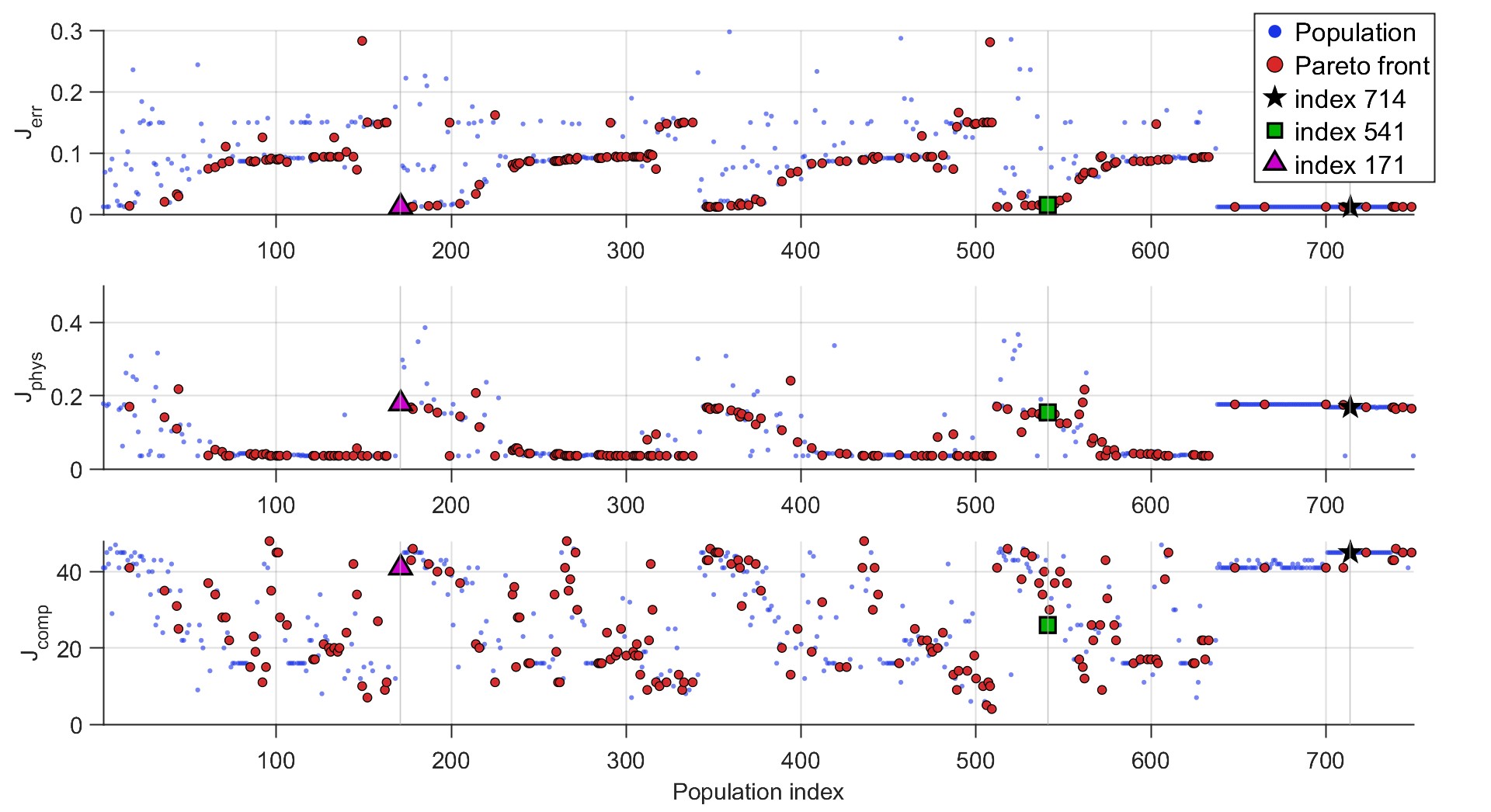}}
  \caption{Objective function results over indices for Run A.}
  \label{fig:j_vs_indices}
\end{figure*}

\begin{table*}
\caption{\label{tab:candidates}%
Selected symbolic regression candidates on the
\((J_\mathrm{err}, J_\mathrm{comp}, J_\mathrm{phys})\) Pareto front.
Candidate~4 is taken from a different run, since that run used a tighter tolerance
instead of the standard value in the \(C_1\) computation; we re-evaluated its
\(J_\mathrm{phys}\) using the standard settings for a fair comparison.}
\begin{ruledtabular}
\begin{tabular}{l c c D{.}{.}{4} D{.}{.}{3} c l}
Candidate & Run & Index &
\multicolumn{1}{c}{$J_\mathrm{err}$} &
\multicolumn{1}{c}{$J_\mathrm{phys}$} &
$J_\mathrm{comp}$ & Role \\
\hline
1 & A & 714 & 0.0121 & 0.168 & 45 & Best fitness \\
2 & A & 541 & 0.0153 & 0.154 & 26 & Simple \\
3 & A & 171 & 0.0126 & 0.178 & 41 & Trade-off \\
4 & B & 639 & 0.0123 & 0.161 & 45 & Tighter tolerance result \\
Haaland & \multicolumn{1}{c}{--} & \multicolumn{1}{c}{--} & 0.1230 & 0.155 &
\multicolumn{1}{c}{--} & Baseline correlation \\
\end{tabular}
\end{ruledtabular}
\end{table*}

\begin{figure}[!hbt]
  \centering
  \includegraphics[width=\columnwidth]{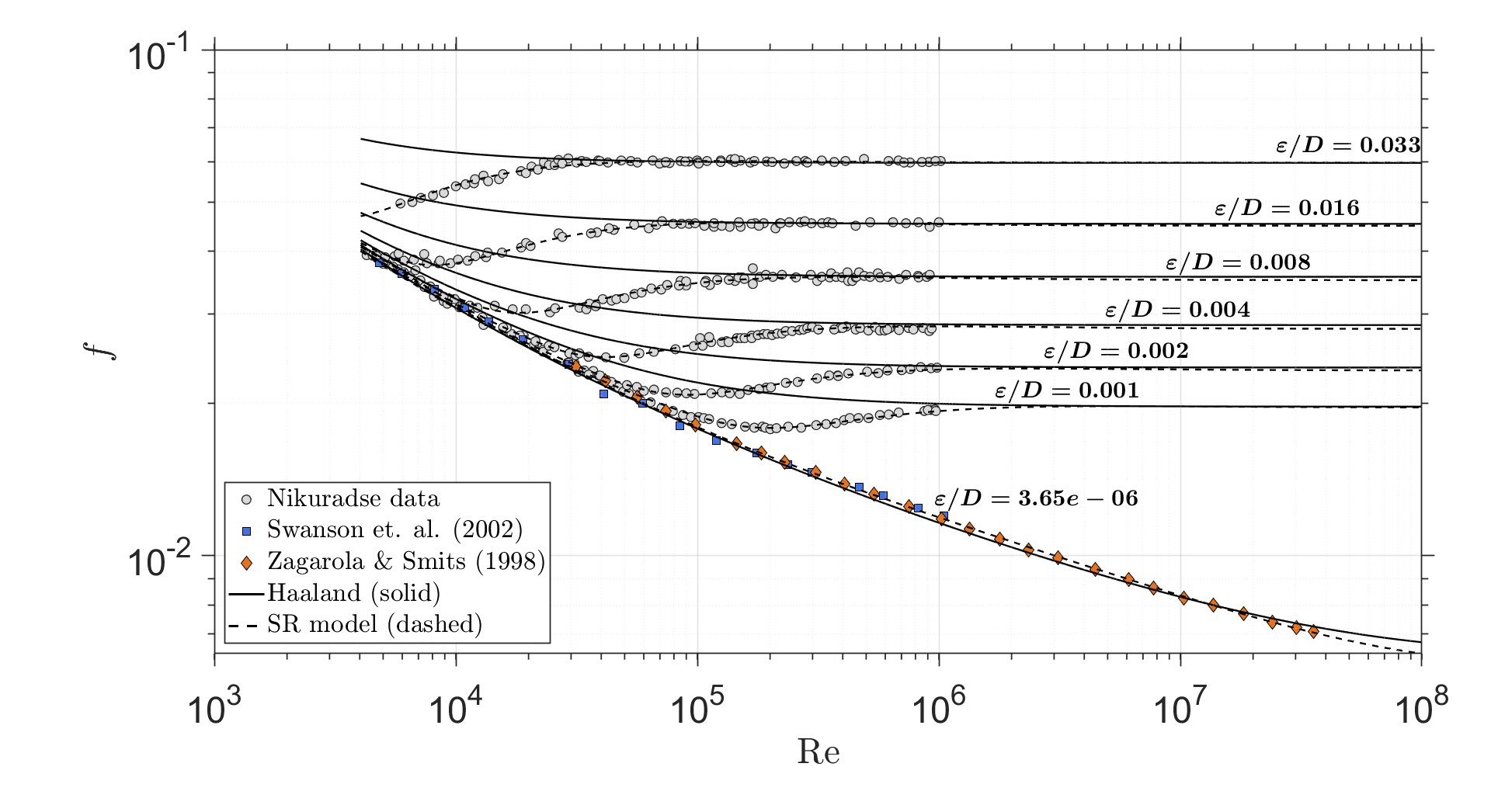}
  \caption{\label{fig:cand1_moody}Nikuradse's friction factor with predetermined dimensionless numbers for Candidate 1 (Re, relative roughness)}
\end{figure}

Candidate~1 (index 714) given in Eq.~\ref{eq:cand1_f} offers the best pure data fit and very good physics compliance, at the cost of being the most complex.  As seen in
Fig.~\ref{fig:cand1_moody}, its dashed curves track the Nikuradse and smooth-pipe data closely across all roughness levels and reproduce the smooth–rough transition with realistic curvature.

\begin{equation}
\label{eq:cand1_f}
\begin{aligned}
f_1
&= 0.2544\,\frac{\varepsilon}{D}
+ 0.1663\left(\frac{\varepsilon}{D}\right)^{\left(\frac{1}{Re\,(\varepsilon/D)^{0.9736}} + 0.9937\right)^{118.2}}
\\[0.2em]
&\quad
+ 0.033\left(
      0.1026\,\frac{\varepsilon}{D}
      - \frac{32840\,\frac{\varepsilon}{D}-2159}{Re}
      \right)^{0.4667}
\\[0.2em]
&\quad
- 0.027\left[
      \frac{\varepsilon}{D}
      + \frac{1}{Re}\left(\frac{1}{Re\,(\varepsilon/D)^{1.1}} + 84.17\right)
      \right]^{-0.029}
\\[0.2em]
&\quad
+ 0.046 \,
\end{aligned}
\end{equation}

Candidate~2 given in Eq.~\ref{eq:cand2_f} represents a different corner of the Pareto front: it is
substantially simpler (26 nodes) and has the best physics scores
(\(J_\mathrm{phys}\approx 0.15\)), but their training errors are about 25\,\% larger than Candidate~1.  In the Moody-style plot, it reproduces the overall structure correctly, but the smooth branch shows a slightly different slope at high \(Re\), and the rough plateau sits a little below the
Nikuradse data for some \(\varepsilon/D\).  These deviations are acceptable from a purely physical standpoint but lead to visibly larger residuals. The correlation of Candidate~2 is:
\begin{equation}
\label{eq:cand2_f}
\begin{aligned}
f_2
&= 0.23\,\frac{\varepsilon}{D}
+ 0.12\left(\frac{\varepsilon}{D}\right)^{\left(\frac{1}{Re\,(\varepsilon/D)^{0.89}} + 1\right)^{220}}
\\[0.2em]
&\quad
+ 0.062\left(
      0.22\,\frac{\varepsilon}{D}
      + \frac{\frac{\varepsilon}{D} + 440}{Re}
      \right)^{0.28}
+ 4.6\times 10^{-3} \,.
\end{aligned}
\end{equation}

\begin{figure}[!hbt]
  \centering
  \includegraphics[width=\columnwidth]{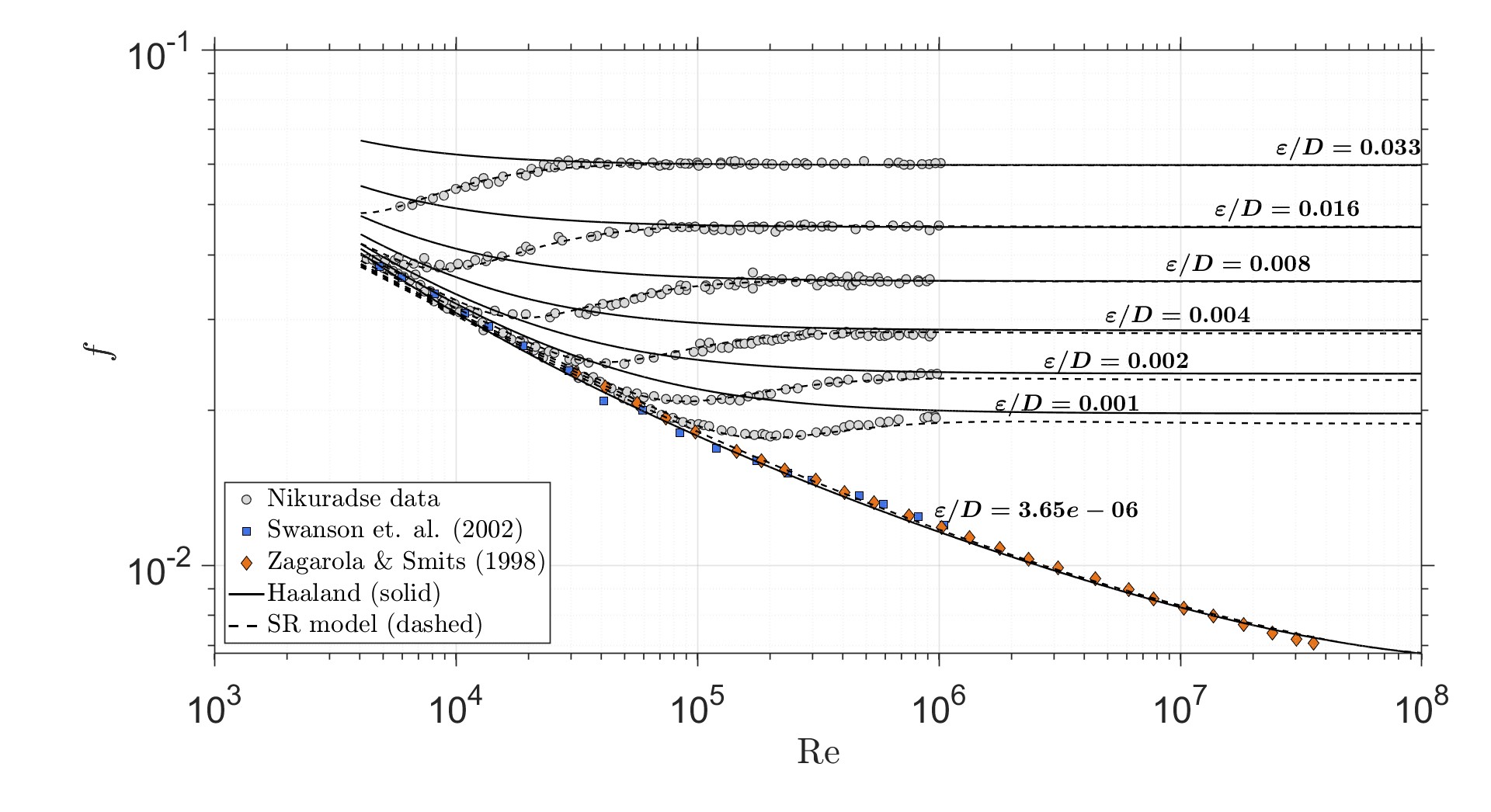}
  \caption{Nikuradse's friction factor with predetermined dimensionless numbers for Candidate 2 (Re, relative roughness).}
  \label{fig:cand2}
\end{figure}

Candidate~3 given in Eq.~\ref{eq:cand3_f} has a similar analytical structure to Candidate 1 (Eq. \eqref{eq:cand1_f}) but with slightly different exponents and numerical constants.  It achieves almost the same error with somewhat lower complexity (41 nodes) at the price of a slightly larger physics score, mainly due to more pronounced deviations in the $C_2$ and $C_3$ envelopes.  On the Moody-style plot, the rough-branch regions are marginally less flat than for Candidate~1, and the transition for
\(\varepsilon/D\approx 10^{-3}\) is shifted. The correlation of Candidate 3 is:
\begin{equation}
\label{eq:cand3_f}
\begin{aligned}
f_3
&= 0.25\,\frac{\varepsilon}{D}
- 0.035\left[
      \frac{\varepsilon}{D}
      + \frac{1}{Re}\left(\frac{1}{(\varepsilon/D)^{1.1}} + 110\right)
      \right]^{-0.029}
\\[0.2em]
&\quad
+ 0.17\left(\frac{\varepsilon}{D}\right)^{\left(\frac{1}{Re\,(\varepsilon/D)^{0.98}} + 0.99\right)^{110}}
\\[0.2em]
&\quad
+ 0.04\left(
      0.0013
      - \frac{2.1\times 10^{4}\,\frac{\varepsilon}{D} - 1.5\times 10^{3}}{Re}
      \right)^{0.48}
\\[0.2em]
&\quad      
+ 0.055 \,
\end{aligned}
\end{equation}

\begin{figure}[!hbt]
  \centering
  \includegraphics[width=\columnwidth]{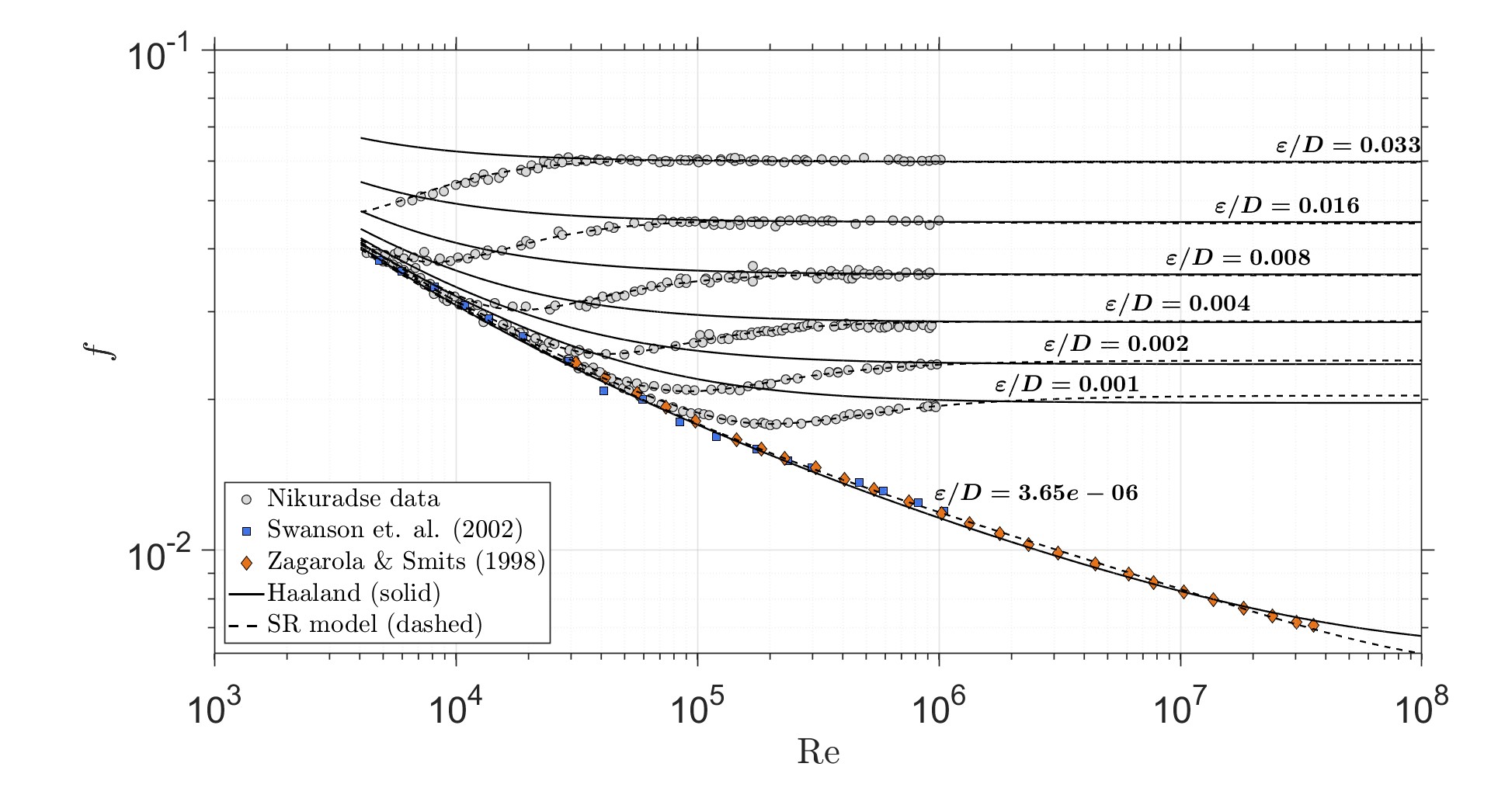}
  \caption{Nikuradse's friction factor with predetermined dimensionless numbers for Candidate 3 (Re, relative roughness).}
  \label{fig:cand3}
\end{figure}
Candidate~4 (from a separate run) given in Eq.~\ref{eq:cand4_f} attains a training error comparable to the best models
(\(J_\mathrm{err}\approx 0.0123\)) with similar complexity (45 nodes), while keeping a competitive
physics score (\(J_\mathrm{phys}\approx 0.161\)).
Its structure differs noticeably from Candidates~1--3: instead of a single dominant
power--law/exponent-gated roughness term, it combines (i) a smooth activation term in \(Re\)
through \(\tanh(32950/Re+1.38)\), (ii) a weak power-law correction in \(\left(\cdot\right)^{0.1584}\),
and (iii) two Reynolds-dependent ``switch'' factors of the form
\((a/x_2)^{c/Re}=\exp\!\big((c/Re)\ln(a/x_2)\big)\).
These exponential-like factors are close to unity at high \(Re\) but can noticeably modulate the prediction at moderate \(Re\), providing additional flexibility in the smooth--rough transition.
The explicit viscous correction \(-27.38/Re\) further helps shape the smooth-branch behaviour.
Overall, Candidate~4 can be viewed as a viable alternative that matches the Nikuradse trends well with similar complexity; we therefore retain it as a strong contender pending further validation.
The correlation of Candidate~4 is:

\raggedbottom
\begin{equation}
\label{eq:cand4_f}
\begin{aligned}
f_4
&= 0.319\,\frac{\varepsilon}{D}
- 0.0345\,\tanh\!\left(\frac{32950}{Re}+1.38\right)
\\[0.2em]
&\quad
+ 0.1316\left[
\left(\frac{\varepsilon}{D}\right)^{2}
- \frac{178.2\,(\varepsilon/D)-22260/Re}{Re}
\right]^{0.1584}
\\[0.2em]
&\quad
- 0.034\left(\frac{0.3391}{\varepsilon/D}\right)^{-11940/Re}
\\[0.2em]
&\quad
+ 0.035\left(\frac{0.1135}{\varepsilon/D}\right)^{-13930/Re}
- \frac{27.38}{Re}
+ 0.03416 \,
\end{aligned}
\end{equation}
\begin{figure}[!hbt]
  \centering
  \includegraphics[width=\columnwidth]{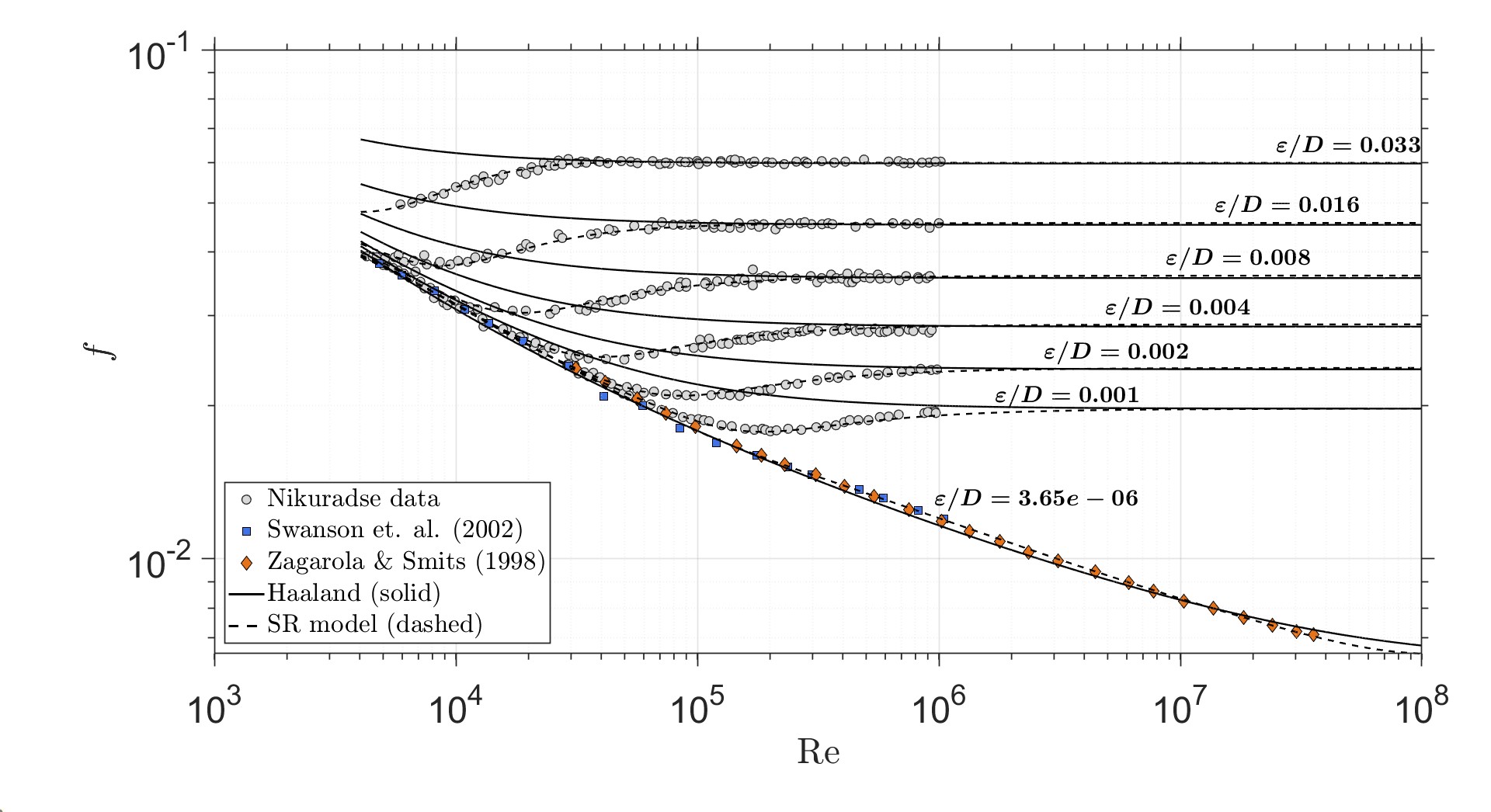}
  \caption{Nikuradse's friction factor with predetermined dimensionless numbers for Candidate 4(Re, relative roughness).}
  \label{fig:cand4}
\end{figure}

After observing the Pareto results, Candidate-1 was selected as the primary model for future analyses.
\subsection{Physical Interpretation of Candidate 1}

For interpretation it is convenient to group the four nonlinear terms in
Candidate 1 (Eq. \eqref{eq:cand1_f}):
\begin{subequations}
\label{eq:cand1_terms}
\begin{align}
T_1 &= 0.2544\,\frac{\varepsilon}{D}
\\[2pt]
T_2 &= 0.1663\left(\frac{\varepsilon}{D}\right)^{\left(\frac{1}{Re\,(\varepsilon/D)^{0.9736}}+0.9937\right)^{118.2}}
\\[2pt]
T_3 &= 0.033\left(0.1026\,\frac{\varepsilon}{D}
           -\frac{32840\,(\varepsilon/D)-2159}{Re}\right)^{0.4667}
\\[2pt]
T_4 &= - 0.027\left[
          \frac{\varepsilon}{D}
          +\frac{1}{Re}\left(\frac{1}{Re\,(\varepsilon/D)^{1.1}}+84.17\right)
       \right]^{-0.029}
\end{align}
\end{subequations}

For clarity, we analyse only the four nonlinear contributions \(T_1\)-\(T_4\); the additive bias in Eq. \eqref{eq:cand1_f} is a constant offset and does not affect the regime structure. 

The term structure is not arbitrary: it mirrors the canonical decomposition of pipe friction into a smooth-wall Reynolds-number dependence, a rough-wall plateau, and a rapid smooth rough crossover familiar from the Moody Diagram, Nikuradse's sand-grain data, and Colebrook-type interpolation. Below we interpret each term in this context, drawing connections to modern high-\(Re\) friction laws and to outer-layer similarity ideas often used to rationalize rough-wall scaling.

\subsubsection{Term \(T_1\): Fully Rough Wall Contribution}

The leading explicit roughness term \(T_1=0.2544\,(\varepsilon/D)\) supplies the dominant fully rough scaling. In the limit \(Re\to\infty\) at fixed \(\varepsilon/D\), the friction factor effectively independent of \(Re\), producing the well-known horizontal plateaux of Moody diagram. This behaviour is experimental signature of the fully rough regime mapped by Nikuradse and later embedded in Colebrook-White: once viscosity no longer controls the near-wall stress, drag is set primarily by relatively roughness. In the perspective of outer-layer similarity, roughness introduces a geometric length scale that replaces the viscous scale in setting the wall stress,
so the leading dependence is on \(\varepsilon/D\) rather than on \(Re\). Candidate~1 enforces this physics explicitly through \(T_1\), yielding (for high Re):
\begin{equation}
f_1 \;\approx\; 0.2544\,\frac{\varepsilon}{D} \;+\; \text{(subdominant corrections)}
\end{equation}

Although classical fully-rough laws are often written in logarithmic form (e.g.~via a roughness function shift in the log law, and thus a log dependence on \(\varepsilon\) when expressed as \(1/\sqrt{f}\)), over the moderate roughness range represented in Nikuradse/Superpipe-style datasets an approximately linear leading-order dependence of \(f\) on \(\varepsilon/D\) can act as an effective surrogate; here \(T_1\) provides precisely that leading roughness sensitivity.

\subsubsection{Term \(T_2\): Roughness Onset Gate}

The most distinctive feature of Candidate 1 is the highly nonlinear term \(T_2\), which activates roughness dependence only beyond a threshold in \((Re,\varepsilon/D)\). By defining

\begin{equation}
\label{eq:switch_mechanism_cand1}
z(Re,\varepsilon/D)
=\left(0.9937+\frac{1}{Re(\varepsilon/D)^{0.9736}}\right)^{118.2},
\end{equation}

We have \(T_2 = 0.1663\,(\varepsilon/D)^{z}\). This makes clear that the large exponent \(118.2\) does not represent a new physical power law; it acts as a steepness parameter for a smooth blending function. Mathematically, a large outer power turns a modest change in the base into an abrupt change in \(z\) so \(T_2\) behaves like a quasi-step activation (a "soft Heaviside"), analogous in effect to using a logistic or \(tanh\)-type switch with a large slope to blend between smooth and rough asymptotes. Candidate 1 makes the switching mechanism explicit in an algebraic form.

Because the outer exponent \(118.2\) makes \(z\) extremely sensitive when the base crosses unity, we define an onset Reynolds number '\(Re_{\mathrm{on}}\)' and set the base in \eqref{eq:switch_mechanism_cand1} to unity, giving 

\begin{equation}
\label{eq:Re_on_cand1}
Re_{\mathrm{on}}(\varepsilon/D)
\;\approx\;
\frac{1}{(1-0.9937)\,(\varepsilon/D)^{0.9736}}
\;\approx\;
\frac{1.59\times10^{2}}{(\varepsilon/D)^{0.9736}}
\end{equation}

For \(Re \lesssim Re_{\mathrm{on}}\), the base exceeds unity so \(z\) grows rapidly; since \(\varepsilon/D<1\), \((\varepsilon/D)^z\) collapses and \(T_2\) is effectively inactive, consistent with a hydraulically smooth response in which roughness elements are submerged within the viscous-affected near-wall region. For \(Re \gg Re_{\mathrm{on}}\), the base approaches \(0.9937\) and \(z\) saturates to a moderate value \(z\approx (0.9937)^{118.2}\approx 0.47\), so \((\varepsilon/D)^z\) is no longer vanishingly small and \(T_2\) contributes at \(O(\varepsilon/D^{0.47})\). In this sense, \(T_2\) provides an explicit and interpretable analogue of the smooth to rough crossover embedded in Colebrook-type formulas: it introduces a sharp transition band in \(Re\) whose location scales approximately as \(Re_{\mathrm{on}}\propto (\varepsilon/D)^{-1}\), consistent with the classical notion that transition is governed by a roughness Reynolds Number (roughness size measured in wall units) reaching \(O(1)\) to \(O(10^2)\). 

\subsubsection{Term \(T_3\): Smooth Wall Reynolds Number Dependence}

The term \(T_3\) carries the principal Reynolds-number dependence of Candidate 1 when \(\varepsilon/D\) is small (hydraulically smooth) or when roughness has not yet been activated by \(T_2\). In the limit \(\varepsilon/D\to 0\), \(T_1\to 0\) and \(T_2\) is strongly suppressed, so the model reduces primarily to the \(Re\)-dependent structure in \(T_3\) (with a weak correction from \(T_4\)).
The explicit appearance of \(1/Re\) inside the base raised to a fractional power (\(0.4667\)) gives \(T_3\) enough flexibility to reproduce the observed curvature of the smooth friction curve: a relatively rapid decrease at moderate \(Re\) that progressively weakens at higher \(Re\) as the flow approaches the logarithmic
asymptote.
This interpretation is aligned with high-$Re$ smooth-pipe measurements from the Princeton Superpipe, a facility developed to access very large $Re_D$ and thereby assess friction-factor relations and their associated universal constants over an extended range of Reynolds number. In particular, McKeon \emph{et al.}~\citep{McKeon2005}, using Superpipe friction data up to \(Re_D=3.5\times 10^7\), showed that the constants in Prandtl's universal friction relation are accurate only over a limited Reynolds-number range and become increasingly biased at higher \(Re_D\), motivating revised constants grounded in a logarithmic-overlap description of the mean velocity profile. In Candidate 1, \(T_3\) plays the analogous role: it supplies the smooth-wall backbone that captures finite-\(Re\) curvature across the combined Nikuradse/Superpipe parameter range before roughness contributions (\(T_1\)--\(T_2\)) become active.

\subsubsection{Term \(T_4\): Weak Interaction Correction}

The remaining term \(T_4\) is deliberately weak: its exponent is \(-0.029\), so \[x^{-0.029}=\exp\!\big(-0.029\ln x\big)\approx 1-0.029\ln x,\] i.e.~\(T_4\) provides a near-logarithmic correction over wide ranges of \(x\). Accordingly, \(T_4\) does not determine the leading asymptotic behaviour; instead it fine-tunes the intermediate curvature and the quantitative matching between the smooth-wall trend (dominated by \(T_3\)) and the roughness-controlled plateau (anchored by \(T_1\), with transition shaped by \(T_2\)). The mixed dependence in
the bracket of \(T_4\) combines \(\varepsilon/D\) and \(1/Re\) in a way that naturally produces small, smooth shifts across the transition without introducing non-physical inflections. From an engineering-correlation perspective, \(T_4\) plays the role of a gentle "shape corrector" that can adjust the depth/curvature of the Colebrook-like shape transition while preserving the correct limits as \(Re\to\infty\) and/or \(\varepsilon/D\to 0\). 

\subsubsection{Synthesis and Implications}

Candidate 1 can therefore be read as an explicit composite correlation that recovers the standard three-part structure of turbulent pipe friction. Term \(T_3\) provides a Superpipe-consistent smooth-wall dependence on \(Re\); term \(T_1\) anchors an \(Re\)-independent roughness-controlled plateau; term \(T_2\) supplies a sharp but continuous activation of roughness effects, mathematically analogous to a steep logistic/\(tanh\) blending function and consistent with the narrow smooth--rough crossover observed in Nikuradse data;
and term \(T_4\) adds a weak, nearly log-like interaction correction that improves quantitative curvature without altering the asymptotic regime structure. This decomposition addresses the common interpretability concern associated with large exponents and small coefficients: the large exponent in \(T_2\) primarily controls switch sharpness, while the small exponent in \(T_4\) produces a gentle corrective adjustment. Taken together, the four terms constitute a compact explicit representation of the classical picture encoded implicitly by Colebrook--White and refined by modern high-\(Re\) measurements: smooth-wall decay with increasing \(Re\), an abrupt transition band whose location scales approximately as \((\varepsilon/D)^{-1}\), and an \(Re\)-independent fully rough tail governed primarily by \(\varepsilon/D\).

\subsection{Constraint behaviour of Candidate~1}

\begin{figure*}
  \centering

  \begin{subfigure}{0.49\textwidth}
    \centering
    \includegraphics[width=\linewidth]{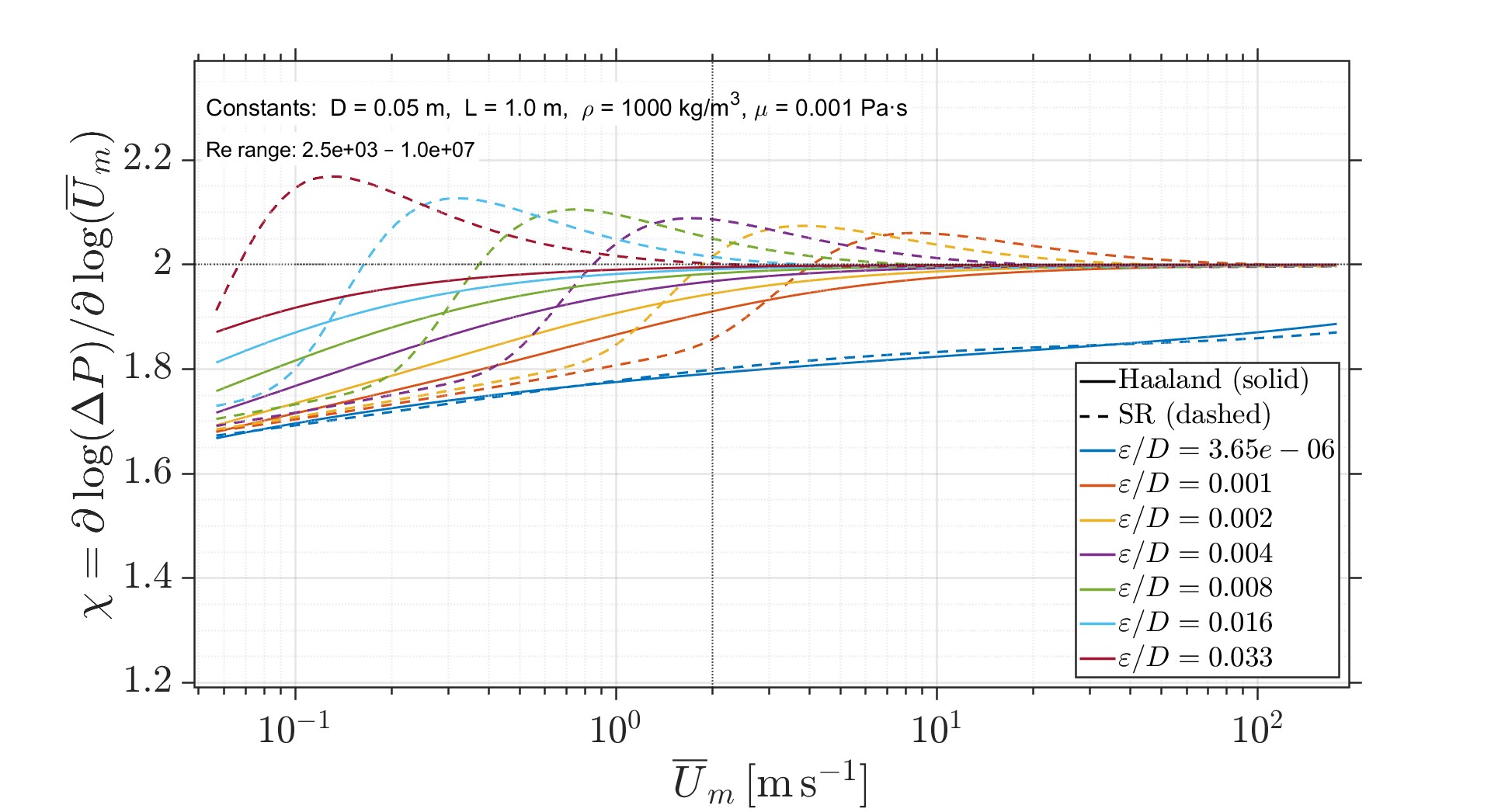}
    \caption{$C_1$ velocity exponent $\chi$.}
    \label{fig:cand1_C1}
  \end{subfigure}\hfill
  \begin{subfigure}{0.49\textwidth}
    \centering
    \includegraphics[width=\linewidth]{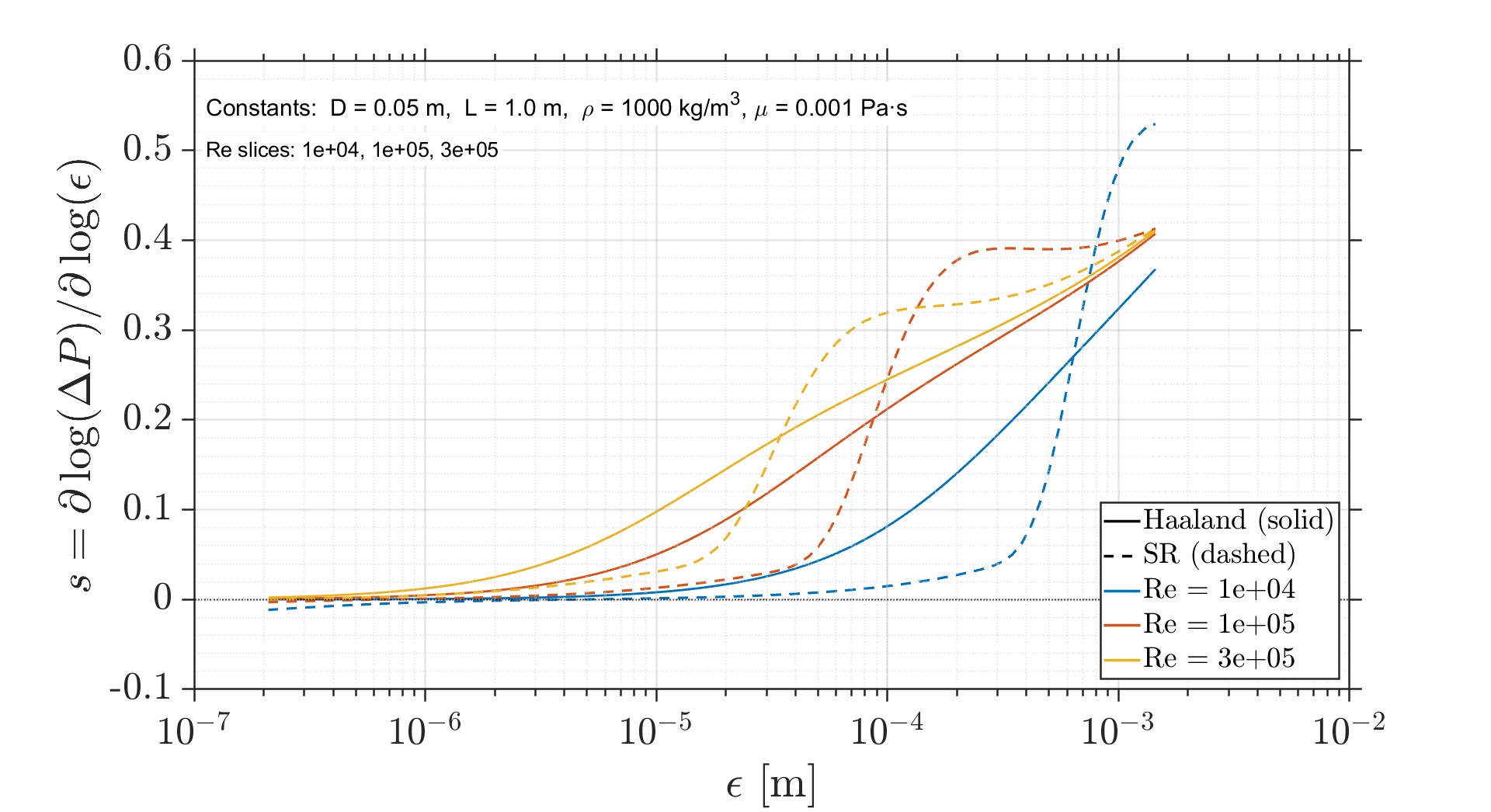}
    \caption{$C_2$ roughness exponent $s$.}
    \label{fig:cand1_C2}
  \end{subfigure}

  \vspace{0.6em}

  \begin{subfigure}{0.49\textwidth}
    \centering
    \includegraphics[width=\linewidth]{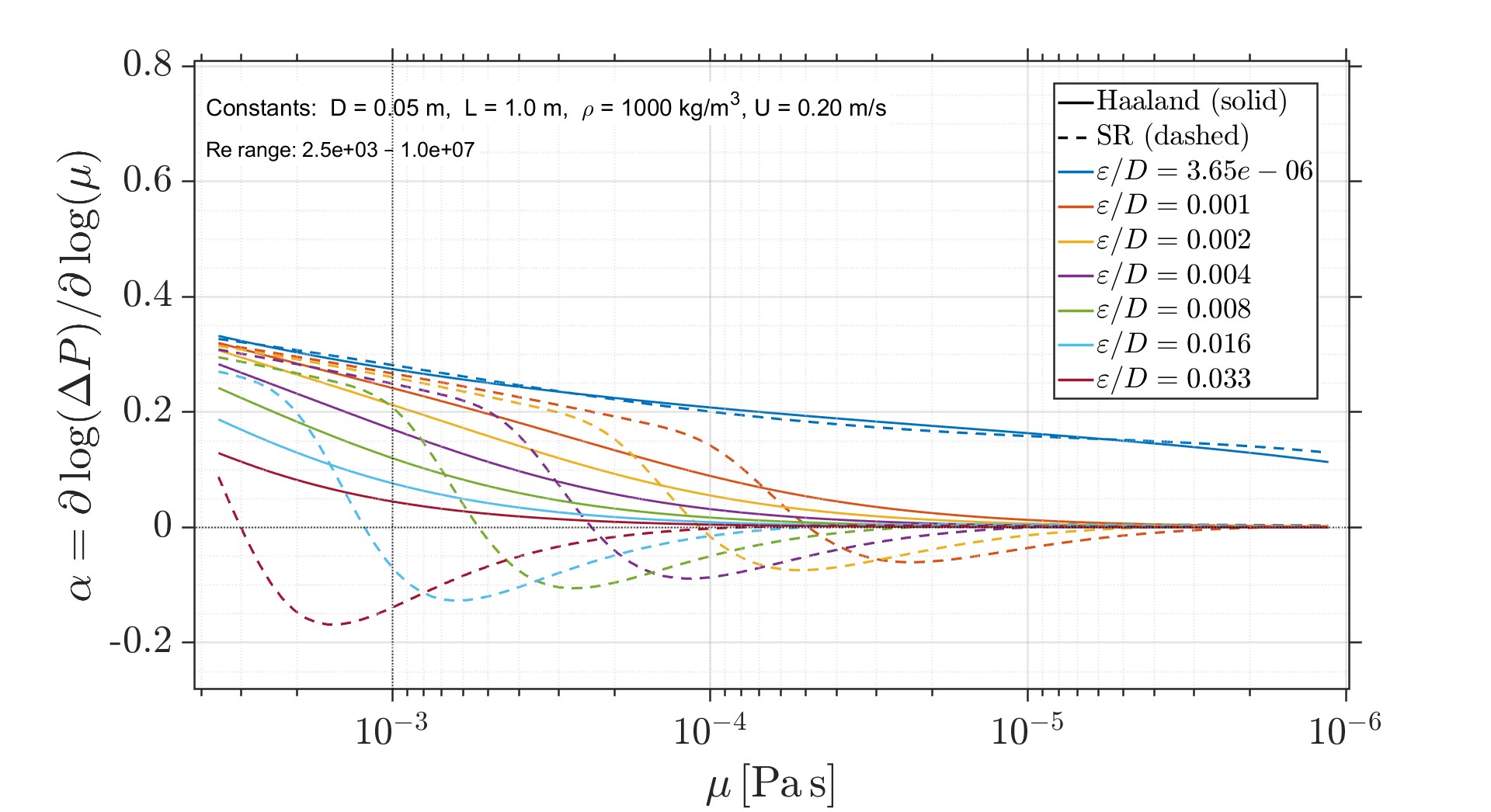}
    \caption{$C_3$ viscosity exponent $\alpha$.}
    \label{fig:cand1_C3}
  \end{subfigure}\hfill
  \begin{subfigure}{0.49\textwidth}
    \centering
    \includegraphics[width=\linewidth]{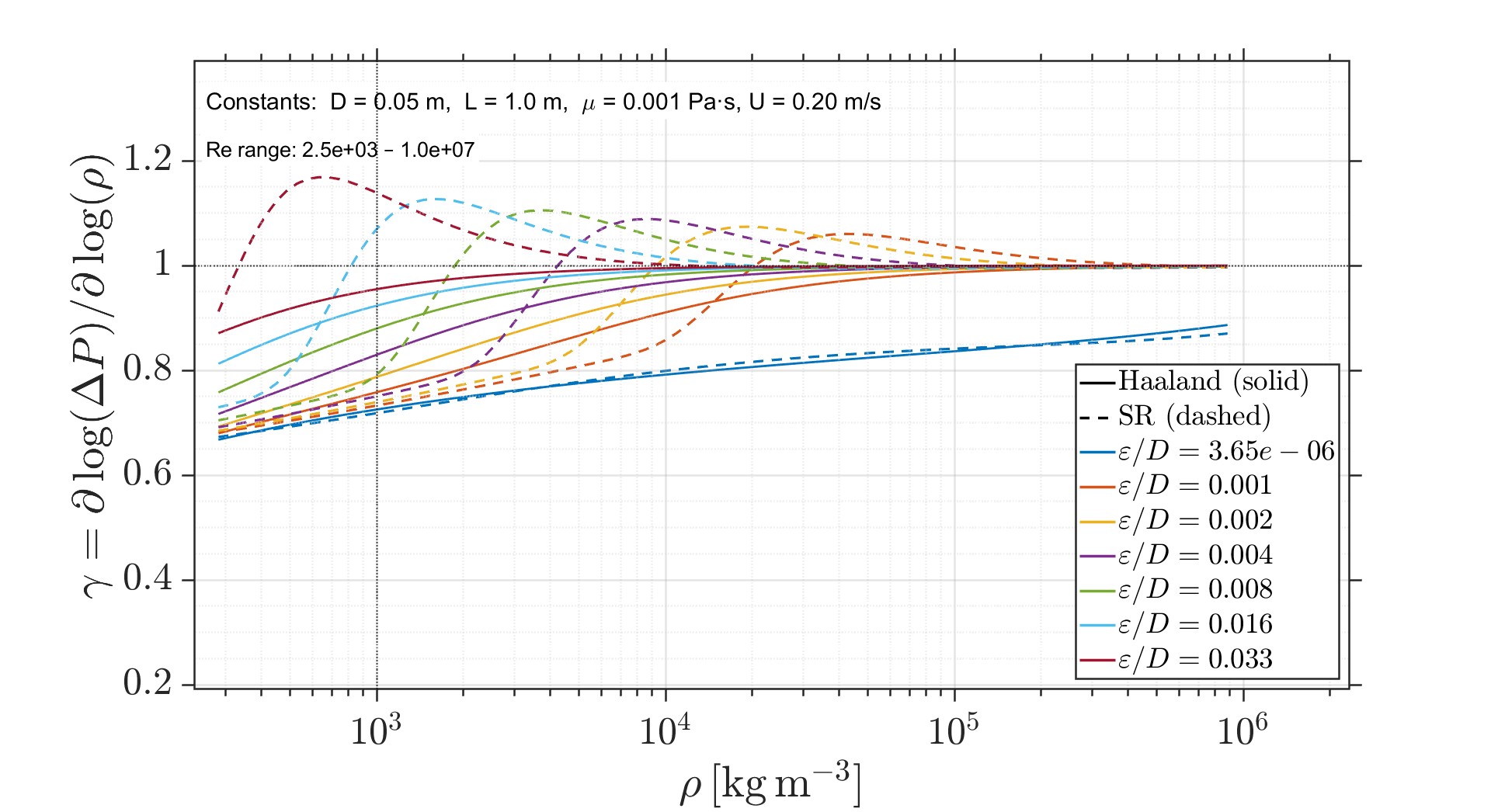}
    \caption{$C_4$ density exponent $\gamma$.}
    \label{fig:cand1_C4}
  \end{subfigure}

  \caption{Constraint checks for Candidate~1 (Eq.~\eqref{eq:cand1_f}), evaluated over the same parameter ranges as the OMA reference. Panels (a)--(d) correspond to $C_1$--$C_4$, respectively.}
  \label{fig:cand1_constraints}
\end{figure*}

Figures~\ref{fig:cand1_C1}--\ref{fig:cand1_C4} show the four constraint checks
for Candidate~1 (Eq.~\eqref{eq:cand1_f}), evaluated in the same parameter ranges
as for the OMA reference. Here \(\chi\), \(s\), \(\alpha\), and \(\gamma\) denote
the local logarithmic exponents of \(\Delta P\) with respect to
\(\overline U_m\), \(\varepsilon\), \(\mu\), and \(\rho\), respectively.

\par\medskip\noindent\textbf{$C_1$ (velocity exponent \(\chi\)).}\quad
In Fig.~\ref{fig:cand1_C1}, the dashed curves for Candidate~1 stay close to the
OMA envelope: for all tested roughness values and across
\(Re\in[2.5\times10^3,10^6]\), the velocity exponent lies between about
\(1.7\) and \(2.2\) and approaches \(2\) toward the high-\(Re\) end. This
behaviour follows directly from Eq.~\eqref{eq:cand1_f}: in the smooth regime the
residual \(1/Re\)-like dependence inherited from \(T_3\) produces an effective
exponent slightly below \(2\), whereas in the rough regime the nearly constant
plateau \(f\approx 0.25\,\varepsilon/D\) forces
\(\Delta P\propto \rho \overline U_m^2\) so that \(\chi\to 2\). The slight
increase in \(\chi\) near the smooth--rough transition indicates finite-\(Re\)
interaction between \(T_2\) and \(T_3\). However, the curves remain within the
OMA-based admissible band, resulting in a low $C_1$ score.

\par\medskip\noindent\textbf{$C_2$ (roughness exponent \(s\) at fixed \(Re\)).}\quad
Figure~\ref{fig:cand1_C2} shows
\(s = \partial\log \Delta P / \partial\log \varepsilon\) at Reynolds-number
slices in the turbulent regime. For each slice, the dashed curves increase
monotonically with \(\varepsilon\) and remain non-negative, so the model
respects the requirement that \(\Delta P\) should not decrease when the wall
becomes rougher. At very small \(\varepsilon/D\) the exponent \(s\) stays close
to zero, consistent with the dominance of the smooth \(1/Re\) term in
Eq.~\eqref{eq:cand1_f}. As \(\varepsilon/D\) enters the transition range, the
activation of \(T_2\) and the growing linear term \(T_1\) cause \(s\) to rise
toward values around \(0.3\mbox{--}0.4\), matching the OMA envelope and the
curvature of the Nikuradse rough branches. The steeper rise of the dashed
curves relative to Haaland reflects that Eq.~\eqref{eq:cand1_f} contains an
explicit \(\varepsilon/D\) prefactor (through \(T_1\)), whereas the Haaland
correlation embeds roughness more implicitly in the logarithm.

\par\medskip\noindent\textbf{$C_3$ (viscosity exponent \(\alpha\) at fixed \(\overline U_m\)).}\quad
In Fig.~\ref{fig:cand1_C3} the dashed curves show how
\(\alpha=\mathrm d\log\Delta P/\mathrm d\log\mu\) varies with \(\mu\) for several
roughness levels at a fixed bulk velocity. For small \(\varepsilon/D\) and
moderate \(Re\) the model behaves as expected from the OMA decomposition:
\(\alpha\approx 0.3\), reflecting a mix of viscous and turbulent contributions.
As \(\mu\) decreases (so \(Re\) increases), the combinations
\(1/(Re(\varepsilon/D)^p)\) in \(T_2\) and \(T_4\) decay and the rough plateau
term \(T_1\) dominates. In this limit, the friction factor becomes nearly
independent of \(\mu\), and \(\alpha\to 0\). The small negative excursions of
\(\alpha\) visible for the roughest pipes arise when the \(1/Re\)-dependent part
of \(T_3\) overcompensates the fading viscous contribution; these are confined
to high-\(Re\), rough conditions where the OMA model itself predicts a very weak
\(\mu\)-sensitivity.

\par\medskip\noindent\textbf{$C_4$ (density exponent \(\gamma\) at fixed \(\overline U_m\)).}\quad
Finally, Fig.~\ref{fig:cand1_C4} reports
\(\gamma=\mathrm d\log\Delta P/\mathrm d\log\rho\). Because \(f_{(1)}\) depends
on \(\rho\) only through \(Re\), increasing \(\rho\) at fixed
\(\overline U_m,\mu\) both multiplies \(\Delta P\) by \(\rho\) and drives \(Re\)
upward. In the smooth regime the \(1/Re\) contribution counteracts part of the
explicit \(\rho\) factor, leading to \(\gamma<1\); in the fully rough regime the
\(Re\)-dependence is negligible and \(\Delta P\propto \rho \overline U_m^2\), so
\(\gamma\to 1\). The dashed curves in Fig.~\ref{fig:cand1_C4} follow exactly
this pattern: they rise monotonically with \(\rho\), stay positive, and
asymptote to a value close to one, in good agreement with the OMA envelope.
This yields a low $C_4$ score while allowing subtle deviations from Haaland in
the intermediate range where the data exhibit some scatter.

Overall, the constraint curves of Candidate~1 remain close to the OMA-based
envelopes in all four projections while tracking the Nikuradse-like transitions
more faithfully than the Haaland correlation.

The candidate models presented above employ only the four base physical constraints of Section~\ref{subsec:Constraints_for_Regression}, which enforce regime-consistent scaling without restricting the internal mathematical structure. A consequence of this flexibility is that the optimization can produce exponents of order $\mathcal{O}(10)^2$, as seen, e.g., in Eq.~\ref{eq:cand1_f}. Although such values exceed what traditional engineering practice typically considers comfortable, where exponents of order unity are preferred, they arise here to achieve sharp but continuous regime transitions.

Additional constraints based, e.g. on engineering experience, can be incorporated to suppress large exponents. However, a single correlation with fixed exponents cannot be expected to represent vastly different operating conditions, such as high-density turbulent liquid-metal flows versus low-density flows at supercritical pressure and temperature conditions. The aim of this work is therefore to provide a robust framework that can be adapted for specific cases of engineering interest.

\subsection{Validation on rough Superpipe datasets}

A final check of the learned correlation is performed on rough-pipe measurements from the Princeton/Oregon "Superpipe" facility. In addition to the smooth-pipe datasets of Swanson \emph{et al.}\cite{Swanson2002} (Oregon) and Zagarola \emph{et al.} \cite{Zagarola1998} (Princeton), which were already included in the training set, we consider two rough-pipe campaigns that were not used for training: the sand-blasted pipe of \cite{Shockling2006}, and the honed pipe of \cite{Langelandsvik2008}. These four datasets span the effective roughness ratios of

\(\,\varepsilon/D \approx 7.6\times10^{-6}\) (Swanson et al.),
\(3.4\times10^{-6}\) (Zagarola \& Smits),
\(5.9\times10^{-5}\) (Shockling et al.), and
\(6.2\times10^{-5}\) (Langelandsvik et al.),
all at Reynolds numbers up to \(Re\sim\mathcal{}10^7\).
The latter two cases therefore provide an out-of-sample test in roughness: their \(\varepsilon/D\) values are almost an order of magnitude larger than the Nikuradse sand-grain pipes used during training.

\begin{figure*}[!hbt]
  \centering
  \includegraphics[width=\textwidth]{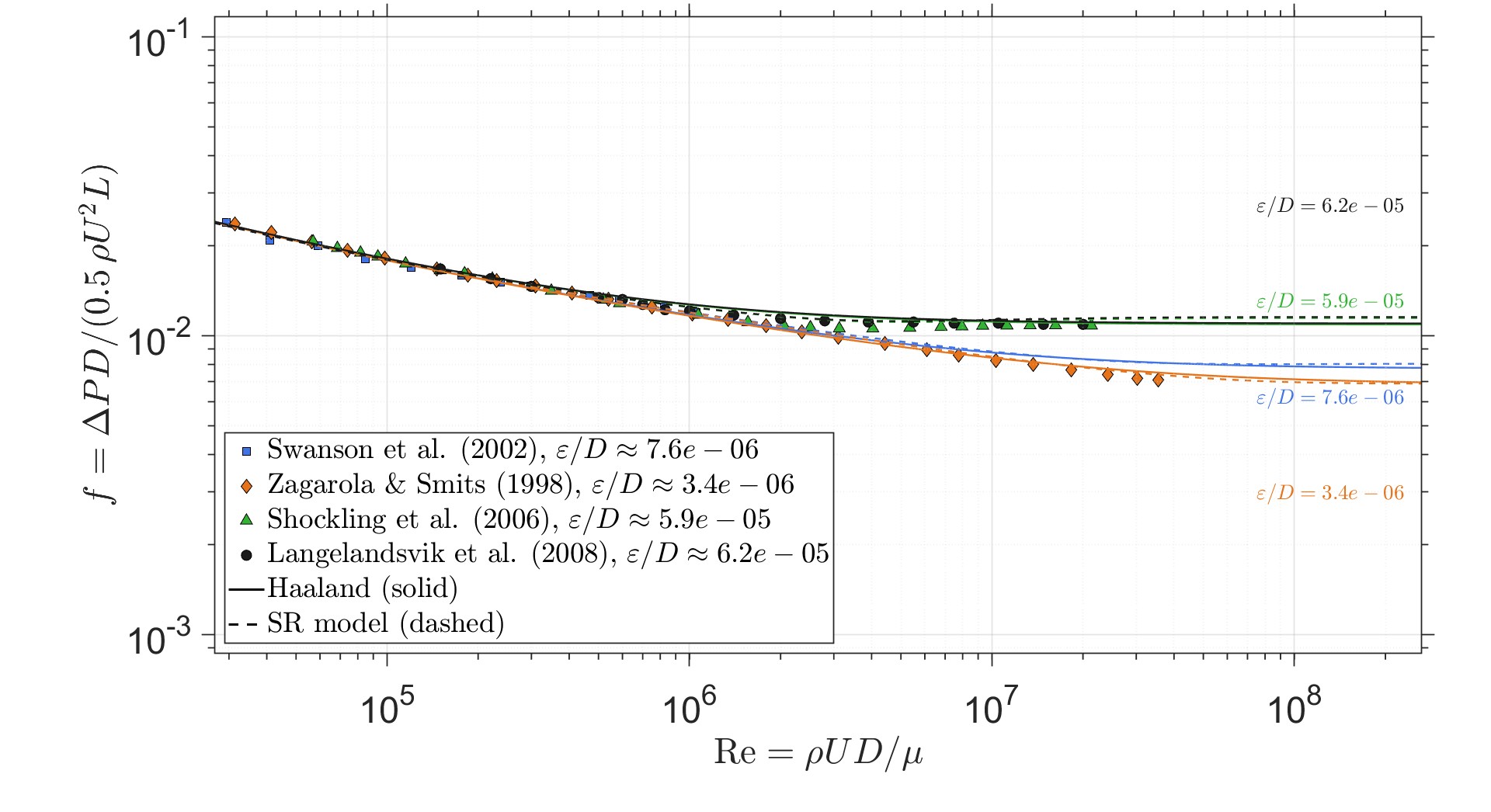}
  \caption{\label{fig:superpipe_validation} Independent high-$Re$ validation of friction-factor predictions for rough pipes. Experimental data (markers) are shown for four nominal $\varepsilon/D$ levels (see legend); the Haaland correlation (solid) and the SR model (dashed) are overlaid. The SR model reproduces the observed trends from $Re\sim10^{5}$ to $10^{8}$ across both low-roughness and fully rough–leaning cases, demonstrating generalization beyond the training set.}
\end{figure*}

Figure~\ref{fig:superpipe_validation} compares these four datasets with the Haaland correlation and with the symbolic regression of Candidate~1. For the two nearly smooth Superpipe smooth Superpipe datasets (Swanson~\citep{Swanson2002} and Zagarola \emph{et al.}~\citep{Zagarola1998}) the SR curve follows the data almost indistinguishably from Haaland across the whole turbulent range, confirming that the learned correlation preserves the classical \(\Delta P \propto Re^{-0.25}\) behaviour in the hydraulically smooth regime and smoothly connects to the laminar-turbulent transition already demonstrated in the Nikuradse-based Moody plot.

For the rough Shockling~\citep{Shockling2006} and Langelandsvik~\citep{Langelandsvik2008} pipes, the situation is more demanding.  Here, the SR model is required to extrapolate in \(\varepsilon/D\) beyond the Nikuradse training range while retaining the correct asymptotic region \(f \sim \text{const}(\varepsilon/D)\) at very large \(Re\).
In Fig.~\ref{fig:superpipe_validation}, the SR prediction (dashed) and Haaland (solid) both capture the overall level and weak Reynolds–number dependence of the data, with the Candidate~1 curve lying between the two rough datasets over most of the range.  Systematic deviations are visible in the fully rough region, where the SR model slightly underestimates the Shockling/Langelandsvik friction factors at the highest Reynolds numbers. This discrepancy is comparable in magnitude to the mismatch between Haaland and the same data, and likely reflects a combination of uncertainties in the effective roughness height of the Superpipe inserts and the known differences between Nikuradse’s uniform sand–grain roughness and the machined surfaces used in the Superpipe facility.
Overall, the Superpipe validation indicates that the physics-informed SR model generalizes reasonably well to roughness levels that were not present in the training set. It preserves the desired smooth-pipe asymptotics and the correct trend toward a roughness-controlled plateau, while exhibiting deviations in the extreme rough regime that are similar to those of the widely used Haaland approximation. This suggests that the combination of dimensional analysis, order-of-magnitude-based constraints, and multi-objective symbolic regression produces a friction-factor correlation that is both interpretable when applied to independent high-Re rough-pipe experiments.

\subsection{Robustness of Symbolic Candidates}

To evaluate how robust each candidate equation is against variations in regression coefficients, we apply a perturbation of \(\pm 1\%\) (uniform, multiplicative noise) and reassess the model response. Here, the goal is instead to probe how sensitive each discovered
functional form is to small changes in its fitted constants---a practical
proxy for the identifiability and numerical conditioning of the learned symbolic mechanisms.

We evaluate robustness of all Candidate equations at three representative \((Re,\varepsilon/D)\) points
chosen to span distinct regimes of the Moody map: a hydraulically smooth case
(small \(\varepsilon/D\)), a fully rough case (large \(\varepsilon/D\) at high
\(Re\)), and an intermediate/transitional case (moderate \(Re\) at appreciable
roughness). This regime-based selection helps distinguish regression candidates that are
globally stable from those that are only locally stable in certain regions of
\((Re,\varepsilon/D)\) space. For each trial, we compute the signed deviation of $J_{\mathrm{err}}$ and $J_{\mathrm{err}}$ relative to the
nominal (unperturbed) prediction at the same operating point. Those deviations are defined as
\begin{eqnarray}
  \Delta J_{\mathrm{err}}=&(f_{\mathrm{dev}}-f_{\mathrm{nom}})/f_{\mathrm{nom}}\\
\Delta J_{\mathrm{phys}}=&J_{\mathrm{phys,dev}}-J_{\mathrm{phys,nom}}
\end{eqnarray}

\begin{figure*}[!hbt]
  \centering
  \begin{subfigure}{0.32\textwidth}
    \includegraphics[width=\linewidth]{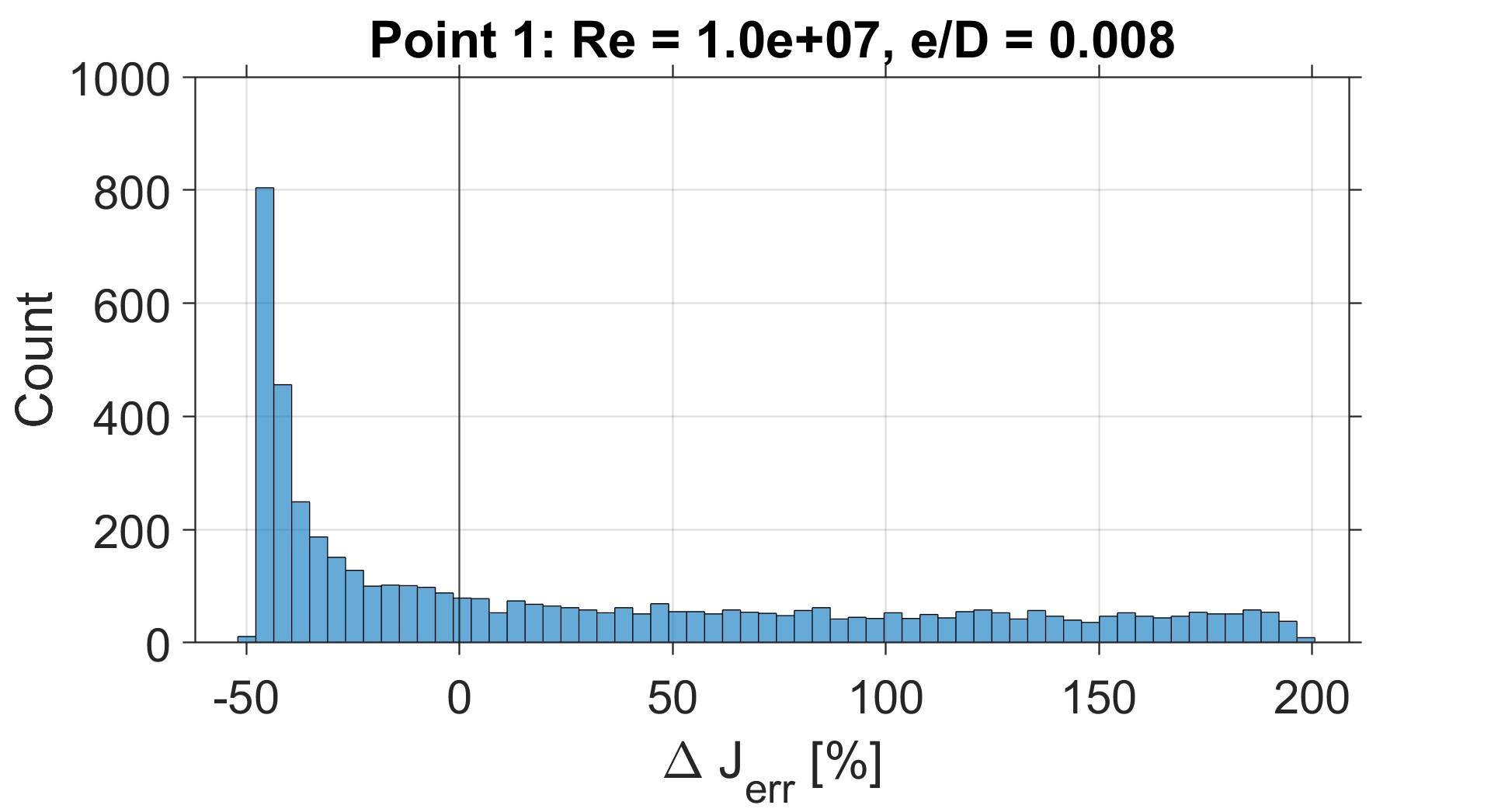}
    \caption{$\Delta J_{\mathrm{err}}$}
  \end{subfigure}\hfill
  \begin{subfigure}{0.32\textwidth}
    \includegraphics[width=\linewidth]{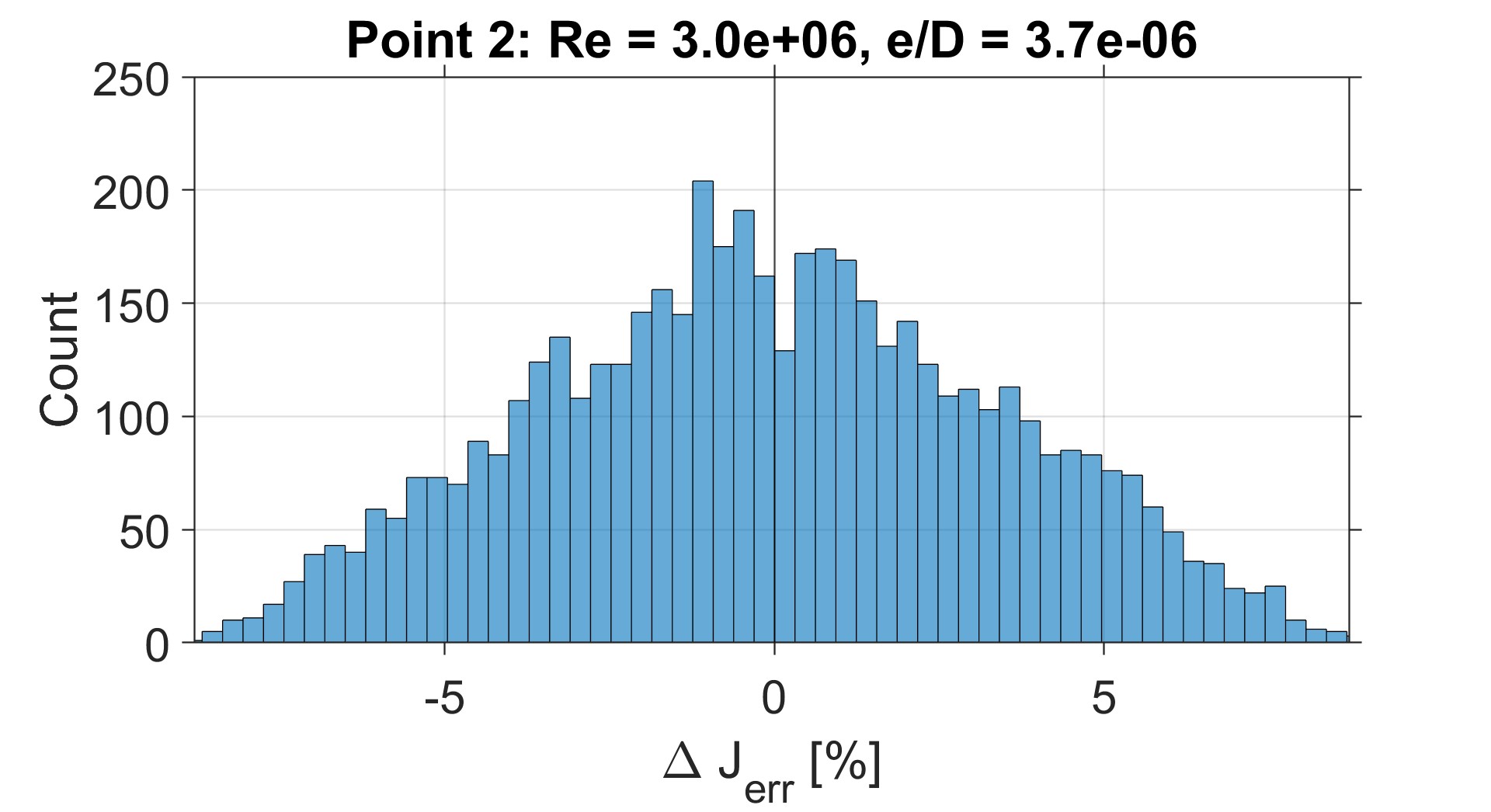}
    \caption{$\Delta J_{\mathrm{err}}$}
  \end{subfigure}\hfill
  \begin{subfigure}{0.32\textwidth}
    \includegraphics[width=\linewidth]{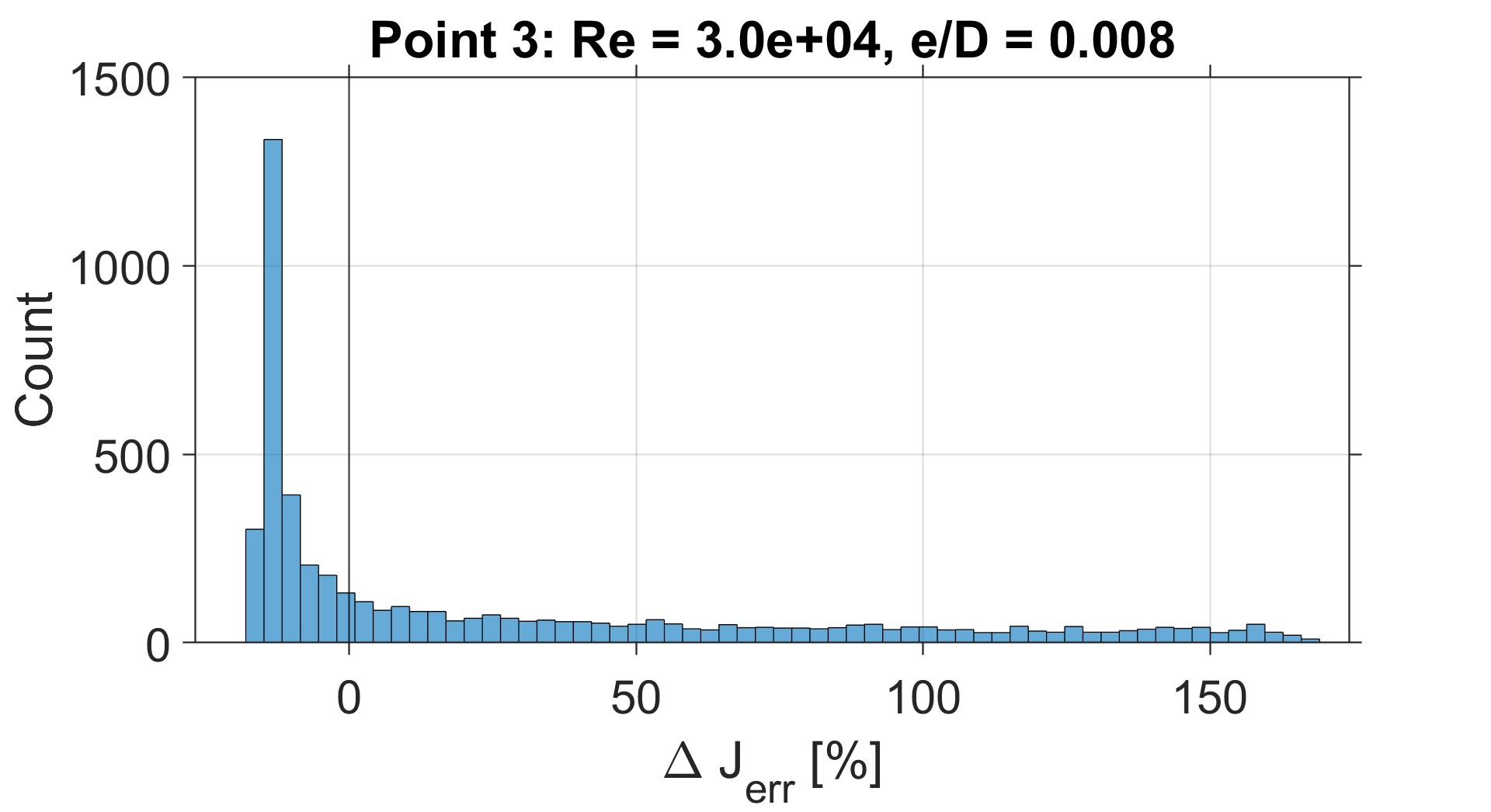}
    \caption{$\Delta J_{\mathrm{err}}$}
  \end{subfigure}

  \vspace{2mm}

  \begin{subfigure}{0.32\textwidth}
    \includegraphics[width=\linewidth]{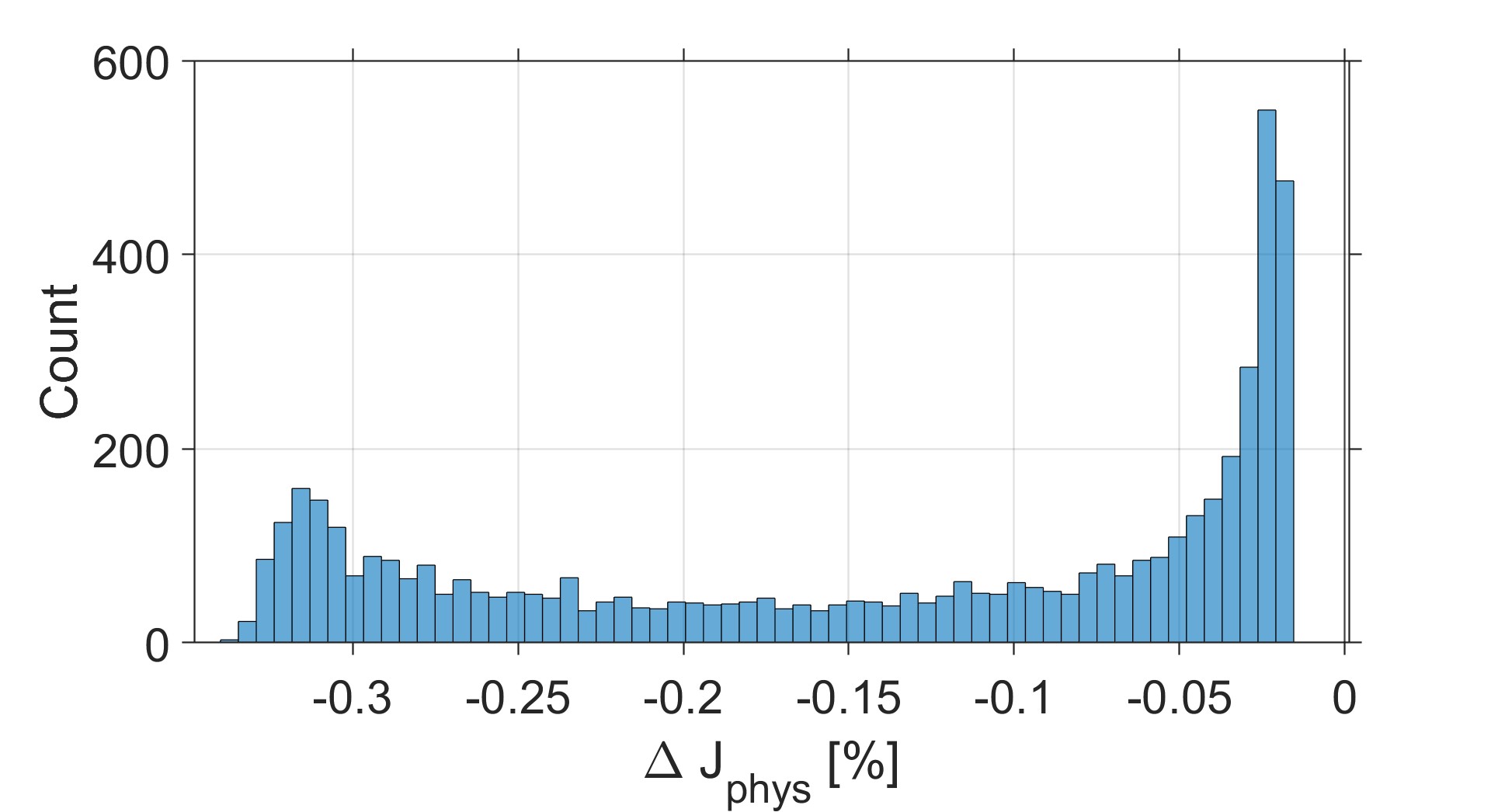}
    \caption{$\Delta J_{\mathrm{phys}}$}
  \end{subfigure}\hfill
  \begin{subfigure}{0.32\textwidth}
    \includegraphics[width=\linewidth]{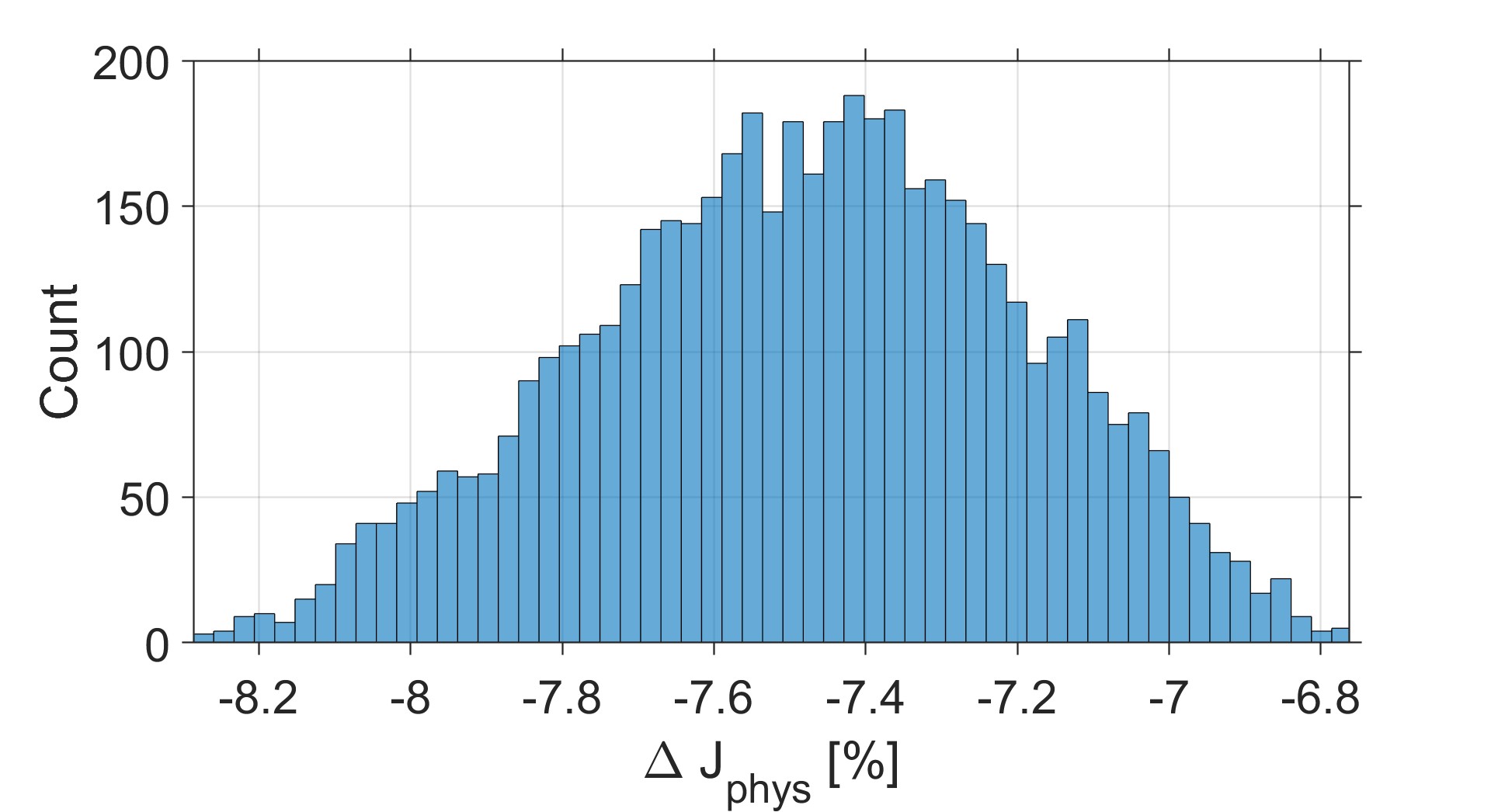}
    \caption{$\Delta J_{\mathrm{phys}}$}
  \end{subfigure}\hfill
  \begin{subfigure}{0.32\textwidth}
    \includegraphics[width=\linewidth]{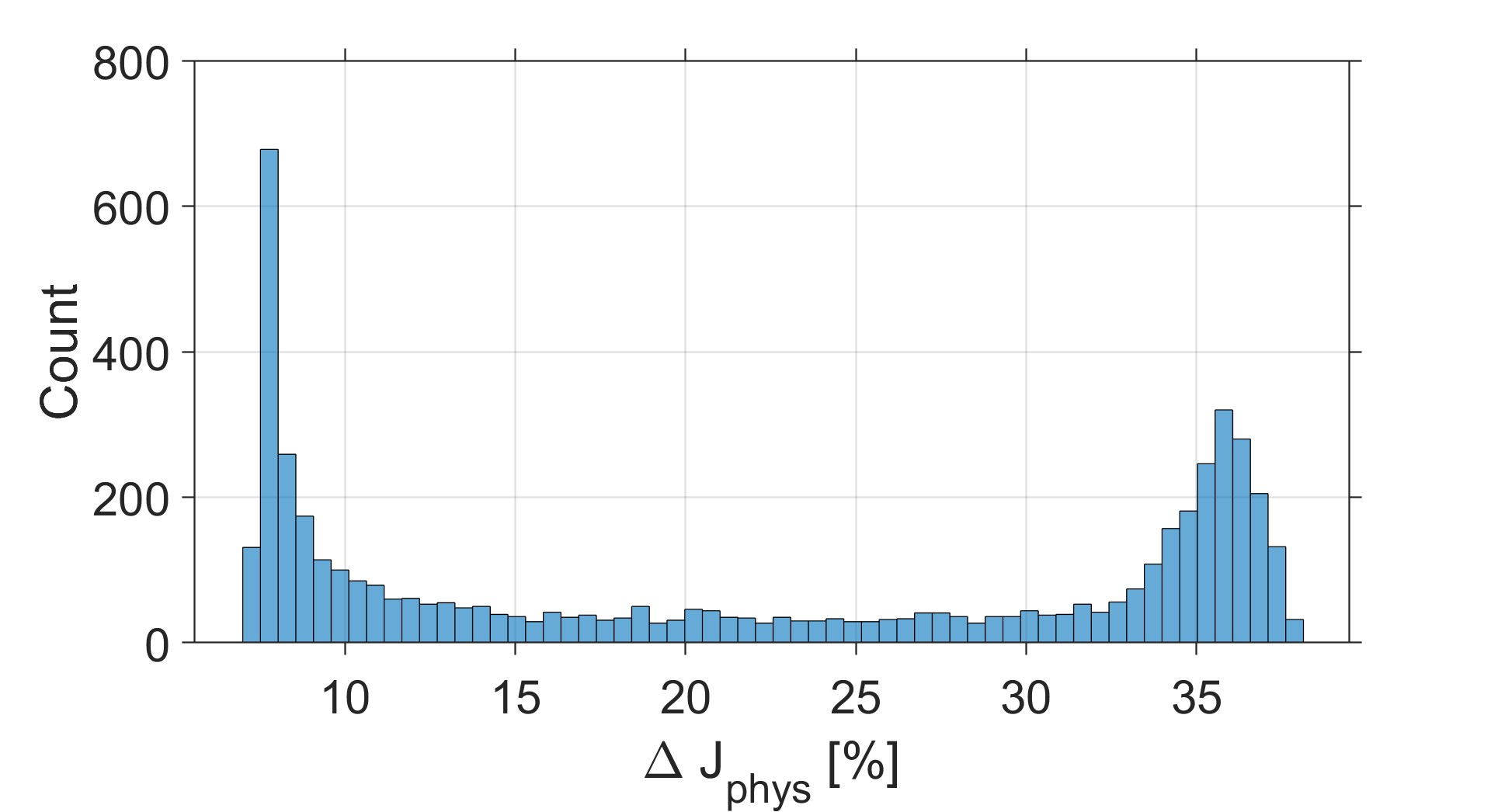}
    \caption{$\Delta J_{\mathrm{phys}}$}
  \end{subfigure}

  \caption{Histograms of the deviation of $J_\mathrm{err}$ and $J_\mathrm{phys}$ from the optimized point corresponding to 1\% deviation of the constants of symbolic regression show the robustness of the equation Candidate 1.}
  \label{fig:robustness_all}
\end{figure*}

The distributions of deviations at three different points are shown in  Fig.~\ref{fig:robustness_all} for Candidate~1. The physics-score deviations are more difficult to interpret directly because
\(J_{\mathrm{phys}}\) is assembled via non-smooth operations (e.g.\ maxima over
constraint components and clipping). As a result, small coefficient changes can
trigger \emph{constraint switching} (changes in the active \(C_k\)) or move a
point across a tolerance boundary, producing asymmetric and sometimes
discontinuous changes in \(J_{\mathrm{phys}}\). This explains why
\(\Delta J_{\mathrm{phys}}\) can appear non-Gaussian or even bimodal in
Fig.~\ref{fig:robustness_all}, and motivates quantile-based summaries rather
than variance-only measures when reporting robustness.

\begin{figure}[!hbt]
  \centering
  \includegraphics[trim={0cm 0cm 0cm 2cm},clip,width=\columnwidth]{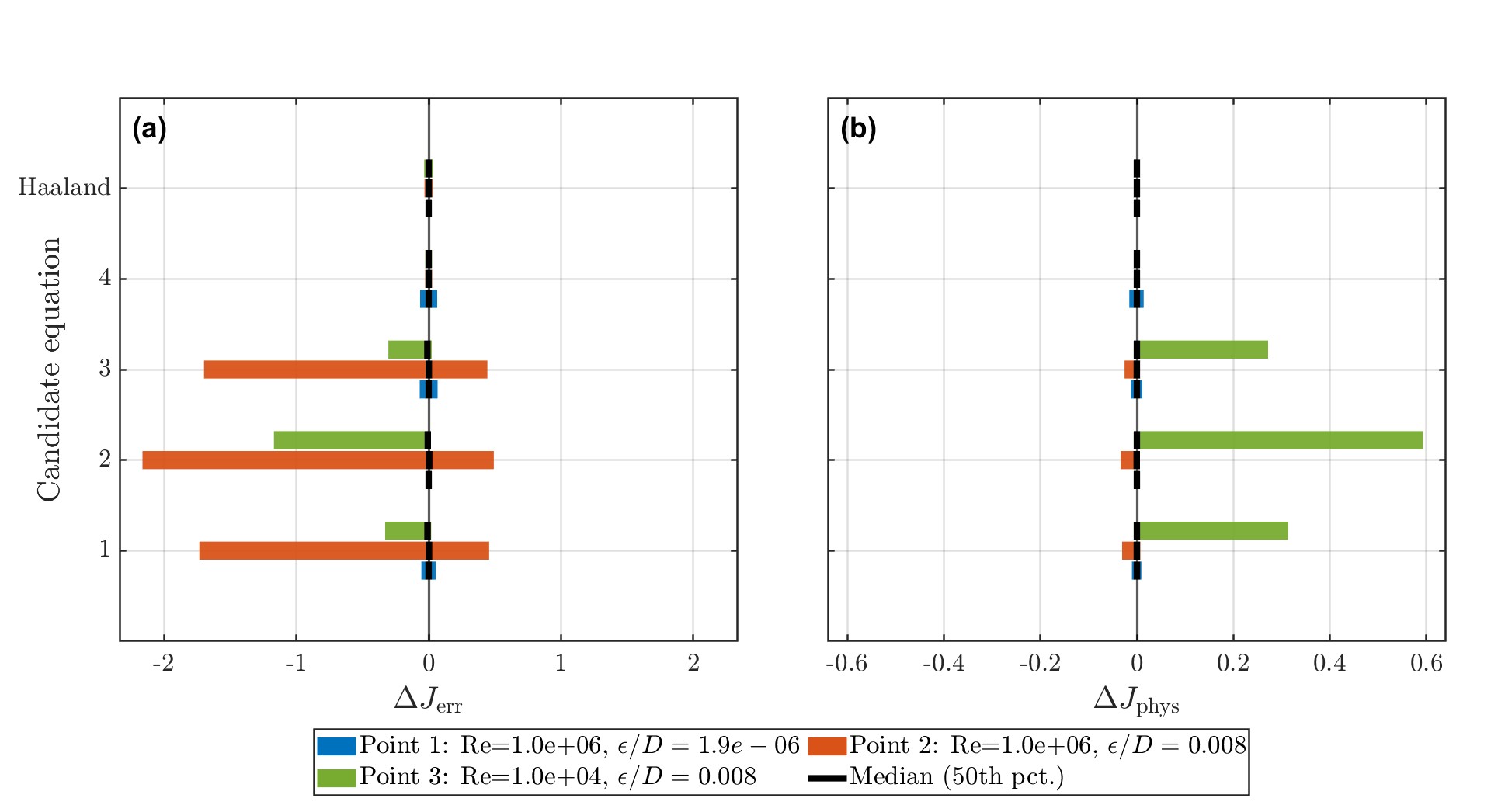}
  \caption{Robustness test on four different candidates in terms of deviation of (a) friction factor (\(\Delta J_{\mathrm{err}}\)) and (b) deviation of constraint score (\(\Delta J_{\mathrm{phys}}\)). The central 95\% interval
of the signed deviations are shown. Black tick marks in those figures indicate the sample median.}
  \label{fig:bar_err}
\end{figure}

We summarize the central 95\% interval
of the signed prediction deviation \(\Delta J_{\mathrm{err}}\)
and the corresponding change in physics score
\(\Delta J_{\mathrm{phys}}\) in Fig.~\ref{fig:bar_err}a and b, respectively. Black tick marks in those figures indicate the sample median. Haaland exhibits a clear deterioration at the transitional point (Point~3),
consistent with the fact that a smooth implicit correlation is not tuned to
local constraint envelopes in that regime. Candidate~4 is comparatively
well-behaved under coefficient noise: its intervals remain narrow at all three
points, and its medians stay close to zero, indicating that its mechanisms
respond smoothly to small constant variations. Candidates~1--3 show markedly
wider tails, i.e.\ occasional large excursions despite medians remaining near
zero. This sensitivity is consistent with sharp ``gating'' substructures (large
outer exponents and threshold-like factors), where small coefficient changes can
shift an internal transition boundary and produce amplified changes at specific
\((Re,\varepsilon/D)\) locations. In that sense, the sensitivity is not purely a
defect: it reflects high confidence in a narrow transition band, while the
median behaviour indicates typical robustness away from rare boundary shifts.

\section{Conclusion}

This work introduced an order-of-magnitude (OMA) and data-based,
physics-informed symbolic regression framework to discover compact friction-factor
correlations across turbulent pipe-flow regimes. By combining classical similarity ideas
\citep{Buckingham1914,Bridgman1922} with multiobjective genetic programming
\citep{Koza1992,Searson2015}, the approach searches directly for explicit algebraic models
that balance predictive accuracy, structural simplicity, and physics-based regime/envelope
requirements.

Unlike conventional single-objective curve fitting and many data-only symbolic regression studies, our formulation (i) enforces physical constraints spanning the smooth, transitionally rough, and fully rough behaviors and (ii) treats model selection as a Pareto decision rather than a single "best-fit" outcome. This perspective is particularly relevant for rough-pipe friction, where the canonical experimental and engineering references highlight distinct regimes and nontrivial transition behavior
\citep{Nikuradse1950,Colebrook1939,Moody1944,Haaland1983,Shockling2006,Langelandsvik2008}.

A central contribution of the study is that the OMA is not used merely as qualitative intuition, but as a constructive model scaffold. Starting from the momentum/energy balance, the pressure drop is decomposed into viscous and turbulence-associated contributions, which motivates a friction-factor structure containing a Reynolds-number-decaying component
and a roughness-controlled component. To capture the transition minimum, we smoothly blend the viscous and turbulent limits using a \(\varepsilon/D, Re\)-dependent switch. This OMA-based approach provides two practical benefits: it yields an interpretable baseline correlation and,
more importantly, it supplies quantitative envelope expectations for sensitivity exponents (\(C_1\)–\(C_4\)), which become lightweight, model-only constraints that guide symbolic regression toward regime-consistent behavior.

A second contribution is a robustness characterization targeting the discovered equations themselves. Rather than perturbing physical inputs, we perturb the fitted regression coefficients to probe numerical conditioning and sensitivity of the symbolic structure—an important practical consideration when models contain sharp nonlinear "gating" mechanisms. The resulting robustness summaries complement regime-wise physical interpretations and help distinguish between models that are accurate but brittle and those that remain stable under small coefficient uncertainty.

Finally, beyond reproducing the training datasets, we validated the correlation learned in  independent rough Superpipe campaigns that were not used for training \citep{Shockling2006,Langelandsvik2008}. These delicate cases provide an out-of-sample test in \(\varepsilon/D\) at very high Reynolds numbers, where the model must extrapolate in roughness while
preserving the correct smooth-pipe asymptotics and the trend toward a roughness-controlled plateau. The symbolic correlation generalizes reasonably to these conditions, with deviations in the extreme rough tail comparable to those of widely used explicit engineering approximations \citep{Haaland1983}. This validation supports the broader claim that combining OMA guidance with multiobjective physics-informed SR can produce correlations that are both interpretable and practically transferable.

Overall, the results show that physics-guided multiobjective symbolic regression can recover interpretable, deployable friction-factor correlations that remain consistent with classical pipe-flow datasets and correlations across regimes, while providing an explicit trade-off surface for accuracy--simplicity--physics compliance. 
In particular, the framework offers a pathway for extending friction-factor predictions to extreme thermal-hydraulic environments, such as liquid metal-cooled nuclear reactors, supercritical water systems, and corroded high-temperature industrial equipment and components, where standard correlations like Haaland's, calibrated primarily from conventional air and water data, may require refinement or replacement. 
The methodology is general and can be ported to other dimensionless-correlation problems where regime structure and asymptotics are known but flexible functional forms are desired.

\section*{Acknowledgments}
This study was supported by the Scientific and Technological Research Council of Türkiye (T\"UB\.{I}TAK), Project No:123M535 and by the EU co-funded partnership CONNECT-NM under Grant Agreement No. 101165375. The authors sincerely thank Prof. Dr. Dr. h.c. Franz Durst and Prof. Dr. El-Sayed Zanoun for the valuable discussion.


\section*{Data Availability Statement}
Collected and combined data and symbolic regression programs are available at GitHub:
\\
\url{https://github.com/emreunal13/GPTIPS2.0_OMAPISR}

\section*{Declaration of Interests}
The authors report no conflict of interest. 

\section*{Author Contribution Statement}
Yunus Emre \"Unal: Conceptualization, Methodology, Software, Investigation, Visualization, Data curation, Validation, Writing -- original draft, Writing -- review \& editing; 

\"Ozg\"ur Ertun\c{c}: Conceptualization, Methodology, Data curation, Supervision, Project administration, Funding acquisition, Writing -- original draft, Writing -- review \& editing; 

\.Ismail Ar\i: Methodology, Validation, Funding, Writing -- review \& editing; 

Ivan Oti\'c: Conceptualization, Methodology, Formal Analysis, Funding, Writing -- review \& editing.






\bibliography{OMASR_REF}

@article{Swanson2002,
   author = {Chris J. Swanson and Brian Julian and Gary G. Ihas and Russell J. Donnelly},
   doi = {10.1017/S0022112002008595},
   issn = {00221120},
   journal = {Journal of Fluid Mechanics},
   month = {6},
   pages = {51-60},
   publisher = {Cambridge University Press},
   title = {Pipe flow measurements over a wide range of Reynolds numbers using liquid helium and various gases},
   volume = {461},
   year = {2002}
}

@article{Smits1998,
   author = {Alexander J. Smits and M. V. Zagarola},
   doi = {10.1063/1.869625},
   issn = {1070-6631},
   issue = {4},
   journal = {Physics of Fluids},
   month = {4},
   pages = {1045-1046},
   publisher = {AIP Publishing},
   title = { Response to “Scaling of the intermediate region in wall-bounded turbulence: The power law” [Phys. Fluids 10 , 1043 (1998)] },
   volume = {10},
   year = {1998}
}

@inbook{Searson2015,
   author = {Dominic P. Searson},
   doi = {10.1007/978-3-319-20883-1_22},
   isbn = {9783319208831},
   booktitle = {Handbook of Genetic Programming Applications},
   month = {1},
   pages = {551-573},
   publisher = {Springer International Publishing},
   title = {GPTIPS 2: An open-source software platform for symbolic data mining},
   year = {2015}
}

@article{Langelandsvik2008,
   author = {L. I. Langelandsvik and G. J. Kunkel and A. J. Smits},
   doi = {10.1017/S0022112007009305},
   issn = {14697645},
   journal = {Journal of Fluid Mechanics},
   month = {1},
   pages = {323-339},
   publisher = {Cambridge University Press},
   title = {Flow in a commercial steel pipe},
   volume = {595},
   year = {2008}
}

@article{Shockling2006,
   author = {M. A. Shockling and J. J. Allen and A. J. Smits},
   doi = {10.1017/S0022112006001467},
   issn = {14697645},
   journal = {Journal of Fluid Mechanics},
   month = {10},
   pages = {267-285},
   publisher = {Cambridge University Press},
   title = {Roughness effects in turbulent pipe flow},
   volume = {564},
   year = {2006}
}

@book{Bridgman1922,
   author = {P W Bridgman},
   city = {New Haven},
   publisher = {Yale University Press},
   title = {Dimensional Analysis},
   year = {1922}
}

@article{Buckingham1914,
   author = {E Buckingham},
   issue = {4},
   journal = {Physical Review},
   pages = {345-376},
   title = {On Physically Similar Systems; Illustrations of the Use of Dimensional Equations},
   volume = {4},
   year = {1914}
}

@article{Blasius1913,
   author = {H Blasius},
   journal = {Forschungsheft des Vereins Deutscher Ingenieure},
   pages = {1-41},
   title = {Das Ähnlichkeitsgesetz bei Reibungsvorgängen in Flüssigkeiten},
   volume = {131},
   year = {1913}
}

@techReport{Nikuradse1950,
   author = {J Nikuradse},
   issue = {Technical Memorandum 1292},
   institution = {National Advisory Committee for Aeronautics},
   note = {English translation of VDI Forschungsheft 361 (1933)},
   title = {Laws of Flow in Rough Pipes},
   year = {1950}
}

@article{Zagarola1998,
   author = {M V Zagarola and A J Smits},
   doi = {10.1017/S0022112098002491},
   journal = {Journal of Fluid Mechanics},
   pages = {33-79},
   title = {Mean-flow scaling of turbulent pipe flow},
   volume = {373},
   year = {1998}
}

@article{Colebrook1939,
   author = {C F Colebrook},
   doi = {10.1680/ijoti.1939.13150},
   issue = {4},
   journal = {Journal of the Institution of Civil Engineers},
   pages = {133-156},
   title = {Turbulent flow in pipes, with particular reference to the transition region between the smooth and rough pipe laws},
   volume = {11},
   year = {1939}
}

@article{Moody1944,
   author = {L F Moody},
   journal = {Transactions of the ASME},
   pages = {671-684},
   title = {Friction Factors for Pipe Flow},
   volume = {66},
   year = {1944}
}

@article{Haaland1983,
   author = {S E Haaland},
   doi = {10.1115/1.3240948},
   issue = {1},
   journal = {Journal of Fluids Engineering},
   pages = {89-90},
   title = {Simple and Explicit Formulas for the Friction Factor in Turbulent Pipe Flow},
   volume = {105},
   year = {1983}
}

@book{White2006,
   author = {Frank M White},
   city = {New York},
   edition = {3},
   publisher = {McGraw–Hill},
   title = {Viscous Fluid Flow},
   year = {2006}
}

@article{Yang2009,
   author = {K S Yang and D D Joseph},
   doi = {10.1080/14685240902791172},
   journal = {Journal of Turbulence},
   pages = {1-26},
   title = {Virtual Nikuradse},
   volume = {10},
   year = {2009}
}

@book{Koza1992,
   author = {John R Koza},
   city = {Cambridge, MA},
   publisher = {MIT Press},
   title = {Genetic Programming: On the Programming of Computers by Means of Natural Selection},
   year = {1992}
}

@article{Schmidt2009,
   author = {Michael Schmidt and Hod Lipson},
   doi = {10.1126/science.1165893},
   issue = {5923},
   journal = {Science},
   pages = {81-85},
   title = {Distilling free-form natural laws from experimental data},
   volume = {324},
   year = {2009}
}

@article{Udrescu2020,
   author = {Silviu-Marian Udrescu and Max Tegmark},
   doi = {10.1126/sciadv.aay2631},
   issue = {16},
   journal = {Science Advances},
   pages = {eaay2631},
   title = {AI Feynman: a physics-inspired method for symbolic regression},
   volume = {6},
   year = {2020}
}

@inproceedings{Cranmer2020,
   author = {Miles Cranmer and Alvaro Sanchez-Gonzalez and Peter Battaglia and Rui Xu and Kyle Cranmer and David Spergel and Shirley Ho},
   booktitle = {Advances in Neural Information Processing Systems},
   pages = {17429-17442},
   title = {Discovering Symbolic Models from Deep Learning with Inductive Biases},
   volume = {33},
   year = {2020}
}

@techReport{Colebrook1937,
   author = {C Colebrook},
   institution = {Imperial College},
   title = {Experiments with Fluid Friction in Roughened Pipes},
   url = {http://royalsocietypublishing.org/rspa/article-pdf/161/906/367/34792/rspa.1937.0150.pdf},
   year = {1937}
}

@techReport{Strickler1981,
   author = {Dr A Strickler and W M Keck},
   institution = {California Institute of Technology},
   title = {Translation T-10 Contributions to the Question of a Velocity Formula and Roughness Data for Streams, Channels and Closed Pipelines},
   year = {1981}
}

@book{Idelchik2005,
   author = {I. E. Idelchik},
   title = {Handbook of Hydraulic Resistance},
   year = {2005},
   ISBN={8179921182,9788179921180},
   publisher = {Jaico Publishing House}
}

@article{McKeon2005,
   author = {B. J. McKeon and M. V. Zagarola and A. J. Smits},
   doi = {10.1017/S0022112005005501},
   issn = {00221120},
   journal = {Journal of Fluid Mechanics},
   month = {9},
   pages = {429-443},
   publisher = {Cambridge University Press},
   title = {A new friction factor relationship for fully developed pipe flow},
   volume = {538},
   year = {2005}
}

\end{document}